\documentclass[3p,12pt]{elsarticle}

\usepackage{indentfirst}
\usepackage{color,graphicx}
\usepackage{subfigure}
\usepackage{threeparttable}
\usepackage{array}
\usepackage{multirow}
\usepackage{amsmath}
\allowdisplaybreaks[4]	
\usepackage{amssymb}
\usepackage{bm}
\usepackage{hyperref}
\usepackage{lineno}

\biboptions{numbers,sort&compress}	

\journal{Journal of Computational Physics}

\begin{document}

\begin{frontmatter}


\title{Adjoint shape optimization from the continuum to free-molecular gas flows}

\author[sustech]{Ruifeng Yuan}

\author[sustech,GHM]{Lei Wu~\corref{cor1}}
\cortext[cor1]{ Corresponding author: wul@sustech.edu.cn (Lei Wu). }

\affiliation[sustech]{organization={Department of Mechanics and Aerospace Engineering, Southern University of Science and Technology},
            city={Shenzhen},
            postcode={518055}, 
            country={China}}

\begin{abstract}
An adjoint-based shape optimization method for solid bodies subjected to both rarefied and continuum gas flows is proposed. The gas-kinetic BGK equation with the diffuse-reflection boundary condition is used to describe the multiscale gas flows. In the vicinity of the gas-solid interface, a body-fitted mesh is utilized, and the sensitivity with respect to the boundary geometry is analyzed through a combined continuous and discrete adjoint methods. The primal and adjoint governing equations are resolved using efficient multiscale numerical schemes, ensuring the precision of the sensitivity analysis in all flow regimes. The sensitivity data is subsequently integrated into a quasi-Newton optimization algorithm to facilitate rapid convergence towards the optimal solution. Numerical experiments reveal that the discretization of the molecular velocity space can induce sensitivity oscillations; however, these can be effectively eliminated by employing appropriate parameterization of the boundary geometry. In optimizing 2D airfoils for drag reduction under varying degrees of gas rarefaction, our method achieves the optimal solution in just a dozen optimization iterations and within a time frame of 5 to 20 minutes (utilizing parallel computation with 40 to 160 cores), thereby underscoring its exceptional performance and efficiency.
\end{abstract}


\end{frontmatter}


\section{Introduction}\label{sec:intro}

Optimizing the shape of objects immersed in fluid flow is a classic problem in modern industry, commonly encountered in the design of vehicles, aircraft, ducts, and more. To address such design problems, the derivative-free surrogate-based methods \cite{forrester2009recent,han2016kriging,bhosekar2018advances} and gradient-based adjoint optimization methods \cite{jameson1988aerodynamic,jameson2003aerodynamic} have been developed. The former is advantageous for its ease of implementation, as it does not require derivative calculations, but it can demand a  large number of optimization iterations particularly when dealing with a large number of design variables. The latter is highly efficient but necessitates the formulation and solution of an adjoint system, which can be a labor-intensive task for engineers, especially when the design problem model is complex.

To date, the majority of optimization methods have been developed for continuum gas flows governed by the Navier-Stokes (NS) equations. However, an increasing number of design challenges now involve rarefied gas flows, especially in cutting-edge fields such as atmospheric re-entry vehicles~\cite{reed2010investigation,li2021kinetic}, vacuum pumps \cite{hablanian1997high,sharipov2005numerical}, lithography \cite{bakshi2009euv}, and nuclear fusion~\cite{tantos2020deterministic}. Since rarefied gas flows cannot be accurately described by the NS equations, there is a need to develop new optimization methodologies.

It is easy to consider incorporating rarefied gas flow solvers into surrogate-based methods due to their fascinating ability of handling black-box problems \cite{lin2023aerodynamic}. This approach is convenient and straightforward to implement, but the computational cost is prohibitive for large-scale applications because rarefied gas flow solvers, whether the direct simulation Monte Carlo (DSMC) method~\cite{Bird1994Molecular} or the discrete velocity method~\cite{Yang1995Rarefied,Mieussens2000DISCRETE}, generally have a much larger computational overhead than NS solvers. Consequently, for design problems involving rarefied gas flows, there is a preference for developing more efficient adjoint-based optimization methods to minimize the number of optimization iterations. 

Recent years have seen the development of adjoint-based topology optimization methods for rarefied gas flow problems~\cite{sato2019topology,guan2023topology,yuan2024design}. These methods can optimize shape and topology simultaneously and have been successfully applied to the design of bend pipes, thermally driven pumps, and airfoils. To manage the topology variation in the flow field, these methods typically optimize the distribution of a material density field that delineates gas and solid regions, leading to two main disadvantages: (i) The mesh is not body-fitted, resulting in first-order accuracy at the gas-solid boundary; (ii) Each mesh cell has a material density value to be optimized, meaning the number of design variables is the same scale of the mesh number. The second disadvantage is particularly undesirable because an excessive number of design variables renders many advanced gradient-based optimizers (e.g., quasi-Newton optimizers) infeasible and slows down the optimization convergence rate. For reference, Sato et al.~\cite{sato2019topology} spent 230 hours on an 80-core computer to complete the design of a 2D thermally driven pump. In the work of Yuan and Wu \cite{yuan2024design}, even with state-of-the-art numerical methods, optimizing 2D airfoils under rarefied conditions took 30-60 minutes with 160-320 cores in parallel, and typically required over 100 optimization steps to achieve the optimum. 

In many industrial design problems, topology variations are not a concern; only shape optimization is necessary. Therefore, for these scenarios, it is essential to develop an adjoint-based shape optimization method with body-fitted mesh, which can offer better boundary accuracy and significantly fewer design variables compared to topology optimization methods.
While there have been some studies on the adjoint analysis of the DSMC \cite{caflisch2021adjoint,guan2023topology} and the gas-kinetic governing equation \cite{sato2019topology,yuan2024design}, to our knowledge, no research has yet focused on sensitivity analysis concerning the shape variation of the gas-solid boundary under the diffuse-reflection condition. 

In this study, we employ the gas-kinetic governing equation, coupled with the diffuse boundary condition, to uniformly describe gas flows from the continuum to free-molecular regimes. We then conduct adjoint analysis to assess the sensitivity with respect to the shape of the gas-solid boundary. By integrating an efficient multiscale numerical solver and a quasi-Newton optimizer, we ultimately propose a shape optimization method that is capable of managing both rarefied and continuum gas flows with exceptional efficiency.

The structure of the paper is as follows. In Section \ref{sec:formula}, we present the foundational theory of the gas-kinetic equation, the formulation of the optimization problem, and the associated adjoint analysis. In Section \ref{sec:method}, we begin with an overview of the multiscale numerical method used to solve the primal and adjoint equations. We then elaborate on the approach to calculating the sensitivity with respect to the shape of the solid boundary and conclude this section by outlining the overall computational procedure of our optimization method. In Section \ref{sec:test}, we conduct numerical tests and provide discussions to validate the method. Finally, Section \ref{sec:conc} offers a summary of the work presented in this paper.

\section{Formulation}\label{sec:formula}

\subsection{Gas-kinetic theory}\label{sec:kinetictheory}

Given that the Navier-Stokes (NS) equation is not sufficiently accurate for rarefied gas flows, we base our method on gas-kinetic theory, which has the capability to describe the dynamics of both continuum and rarefied gas flows in a unified manner~\cite{Cercignanibook1988,Lei2022}.
In the gas-kinetic theory, the state of the gas is described by the molecular velocity distribution function $f$, from which the macroscopic flow variables can be obtained by integration in the molecular velocity space. For example, the conservative variables can be calculated as
\begin{equation}\label{eqn:w_int}
\bm W = \int {\bm \psi fd\Xi },
\end{equation}
where $\bm W=(\rho,\rho\bm u,\rho E)^\top$ is the vector for the densities of mass, momentum and energy. The specific energy is $E={\bm u}^2/2 + {RT}/({\gamma  - 1})$, with $T$ being the gas temperature, $R$ the specific gas constant, and $\gamma$ the specific heat ratio. In the integration of Eq.~\eqref{eqn:w_int}, $\bm \psi$ is the vector of moments $\bm \psi  = {\left( {1,\bm v,\frac{1}{2}{{\bm v}^2}} \right)^\top}$ where $\bm v=(v_1,v_2,v_3)$ is the molecular velocity, and $d\Xi  = dv_1dv_2dv_3$ is the molecular velocity space element. 

The governing equation for the distribution function $f$ is the Boltzmann equation \cite{Cercignanibook1988}, however due to the heavy computation of the collision operator here we consider its model equation, i.e.~the widely used Bhatnagar-Gross-Krook (BGK) equation \cite{bhatnagar1954model}
\begin{equation}\label{eqn:bgk_org}
\frac{{\partial f}}{{\partial t}}{\rm{ + }}\bm v \cdot \nabla f = \frac{{g - f}}{\tau }.
\end{equation}
In this model equation, the equilibrium state $g$ follows the Maxwellian distribution
\begin{equation}
g =g_{\rm M}(\rho,\bm{u},T)= \rho {\left( {\frac{1}{{2\pi RT}}} \right)^{\frac{3}{2}}}{\exp\left( - \frac{{{{(\bm v - \bm u)}^2}}}{{2RT}}\right)}.
\end{equation}
The relaxation time $\tau$ is calculated by
\begin{equation}
\tau  = \frac{\mu}{p},
\end{equation}
where $p=\rho RT$ is the pressure and $\mu$ is the dynamical viscosity. We consider the gas molecules with the hard-sphere model \cite{Bird1994Molecular} therefore it has $\mu  \propto \sqrt T $. 

Take moments of the BGK equation \eqref{eqn:bgk_org} about $\bm \psi$, one can get the transport equation for the macroscopic conserved quantities
\begin{equation}
\frac{{\partial \bm W}}{{\partial t}} + \nabla  \cdot {\bf{F}} = 0,
\end{equation}
where ${\bf{F}} = \int {\bm v\bm \psi fd\Xi } $ is the flux tensor, and the conservation condition $\int {\bm \psi  \frac{{g - f}}{\tau } d\Xi }  = \bm{0}$ is fulfilled for the collision term.

In rarefied gas dynamics, the Knudsen  number (Kn) serves as a dimensionless parameter indicating the level of rarefaction. It is defined as the ratio of the molecular mean free path to the characteristic length $l_{\rm ref}$ of the gas flow.
The mean free path is determined by the hard-sphere model \cite{Bird1994Molecular}, and the Knudsen number can be calculated as
\begin{equation}\label{eqn:kndefine}
{\rm{Kn}} = \frac{{16}}{5}\frac{\tau }{{{l_{{\rm{ref}}}}}}\sqrt {\frac{{RT}}{{2\pi }}} .
\end{equation}
According to the magnitude of Knudsen number, gas flows are usually qualitatively classified into different flow regimes \cite{tsien1946superaerodynamics}. In the continuum flow regime ($\rm{Kn}<0.001$), the gas flow admits the linear constitutive relationship (Newton's law for viscosity and Fourier's law for heat conduction) and can be described by the NS equation. In the slip flow regime ($0.001<\rm{Kn}<0.1$), the velocity slip and temperature jump occur on the gas-solid boundary and the no-slip condition is invalid. In the transition flow regime ($0.1<\rm{Kn}<10$) and the free-molecular flow regime ($\rm{Kn}>10$), the gas can be no longer viewed as the continuum medium and the NS equation is invalid, and the flow should be described by the molecular transport-collision dynamics. 

The Boltzmann kinetic equation is applicable across all flow regimes. In the continuum flow regime, the kinetic equation can be related to the NS equations through the Chapman-Enskog expansion \cite{chapman1990mathematical}. However, in the continuum flow regime, the kinetic equations possess a very stiff collision term, necessitating special numerical treatment. This topic will be further elaborated in Section \ref{sec:method_scheme}.

\subsection{Reduced governing equation}

It should be noted that the BGK equation \eqref{eqn:bgk_org} is formulated in 3D physical space and 3D molecular velocity space. However, the current study focuses on 2D problems. Consequently, a reduction technique  can be employed to significantly reduce memory usage and computational expenses \cite{chu1965kinetic,Yang1995Rarefied}.  The original distribution function $f$ can be reduced to two reduced (or marginal) distribution functions $f_1$ and $f_2$:
\begin{equation}\label{eqn:f_reduce}
\left.
\begin{aligned}
{f_1}(\bm x,\bm v,t) &= \int {f(\bm x,\bm v,v_3 ,t)dv_3 } \\
{f_2}(\bm x,\bm v,t) &= \int {\frac{1}{2}{v_3 ^2}f(\bm x,\bm v,v_3 ,t)dv_3 } 
\end{aligned}
\right\},
\end{equation}
where now $\bm x=(x_1,x_2)$ and $\bm v=(v_1,v_2)$ belong to a space of $2$ degrees of freedom. Then the governing equation for the reduced distribution function $\bm f=(f_1,f_2)^\top$ can be obtained by multiplying Eq.~\eqref{eqn:bgk_org} by $(1,v_3^2/2)^\top$ and integrating with respect to $v_3$:
\begin{equation}\label{eqn:bgk_reduce}
\frac{{\partial \bm f}}{{\partial t}} + \bm v \cdot \nabla \bm f = \frac{{{\bm g } - \bm f}}{{{\tau }}},
\end{equation}
where the equilibrium state $\bm g={\bm g_{\rm{M}}}(\rho ,{\bm u },{T })={({g_{1}},{g_{2}})^\top}$ is
\begin{equation}\label{eqn:maxwell_reduce}
{g_{1}} =\frac{ \rho}{2\pi RT} \exp\left( - \frac{(\bm v - {\bm u})^2}{2R{T}}\right),\quad
{g_{2}} = \frac{{{1 }}}{2}R{T }g_{1},
\end{equation}
and accordingly the macroscopic variables $\bm W$ can be calculated from the reduced $\bm f$ by
\begin{equation}\label{eqn:macint_reduce}
\bm W = \int {\bf{\Psi }}    \cdot  \bm f d\Xi , 
\end{equation}
where $\bf{\Psi }$ is the moments tensor
\begin{equation}
{\bf{\Psi }} = \left( {\begin{array}{*{20}{c}}
\begin{array}{l}
1\\
\bm v\\
{\textstyle{1 \over 2}}{\bm v^2}
\end{array}&\begin{array}{l}
0\\
0\\
1
\end{array}
\end{array}} \right).
\end{equation}
Likewise, hereafter $d\Xi  = dv_1dv_2$ denotes the reduced velocity space element of $2$ dimensions.

\subsection{Boundary condition}\label{sec:boundary}

We consider two types of boundary conditions. 
The first is the Dirichlet boundary condition for the far-field free-stream boundary:
\begin{equation}\label{eqn:bc_drlt}
\bm f = \bm f_{\rm d} \quad {\rm{in}}\quad {\Gamma _{\rm d}} \times {\Xi ^ - }.
\end{equation}
Here $\Gamma _{\rm d}$ denotes the boundary with the Dirichlet condition applied, ${\Xi ^ \pm } = \left\{ \bm v | \bm v \cdot \bm n \gtrless 0 \right\}$ representing the gas molecules flow out of or into the boundary with a normal unit vector $\bm n$ pointing outward from the gas field, and $\bm f_{\rm d}$ is a given fixed velocity distribution for the molecules flow into the computational domain. Normally $\bm f_{\rm d}$ is given as a Maxwellian distribution~\eqref{eqn:maxwell_reduce} with the macroscopic state of the far-field condition.

The second is the diffuse boundary condition imposed on the gas-solid surface. When the gas molecules hit the solid wall ${\Gamma _{\rm w}}$, the velocity distribution of the reflecting molecules follows the Maxwellian distribution:
\begin{equation}\label{eqn:formula_bgk_fdw0}
\left.
\begin{aligned}
\bm f &= {\bm g_{\rm{w}}},\\
{\bm g_{\rm{w}}}&={\bm g_{\rm{M}}}({\rho _{\rm{w}}},{\bm u_{\rm{w}}},T_{\rm{w}}),
\end{aligned}
\right\}
\quad {\rm{in}}\quad {\Gamma _{\rm w}} \times {\Xi ^ - }
\end{equation}
where ${\bm u_{\rm{w}}}$ is a given wall velocity and since we don’t consider the wall motion here it satisfies $\bm u_{\rm{w}} = \bm 0 $; $T_{\rm{w}}$ is a given wall temperature; $\rho _{\rm{w}}$ is determined from the zero mass flux condition at the gas-solid surface as
\begin{equation}\label{eqn:formula_bgk_fdw}
\left.
\begin{aligned}
&{F_{\rho{\rm{w}} }} + \int_{{\Xi ^ - }} {\bm v \cdot \bm n{g_{{\rm{w}},1}}d\Xi }  = 0,\\
&{F_{\rho{\rm{w}}}} = \int_{{\Xi ^ + }} {\bm v \cdot \bm n{f_1}d\Xi } .
\end{aligned}
\right\}
\end{equation}
In the continuum flow regime, the diffuse boundary condition~\eqref{eqn:formula_bgk_fdw0} can automatically recover the no-slip boundary condition, while for the rarefied flow it can recover non-equilibrium effects such as the velocity slip and temperature jump on the gas-solid interface. Therefore, it is applied to handle both rarefied and continuum flow problems.

\subsection{Optimization problem}

We now articulate the optimization problem using mathematical notation. First we declare that only the steady-state problem is considered here, therefore there will be no term associated with the time $t$. We are about to optimize the shape of the solid body immersed in the gas flow domain $\Omega$ which has the far-field boundary $\Gamma _{\rm d}$ and the gas-solid boundary $\Gamma _{\rm w}$. The design variables are a set of parameters determining the shape of the solid boundary $\Gamma _{\rm w}$. This parameterization for the shape of $\Gamma _{\rm w}$ will be detailed later in Section \ref{sec:cst}; before that, let's refer to it as the design variable. 

For the optimization objective, we mainly consider the momentum/heat  transported through the gas-solid boundary, which can be expressed as the integration of a certain moment of the mass flux $\bm v \cdot \bm n f_1$ over the solid boundary $\Gamma _{\rm w}$. Besides, the volume of the solid body, which can be easily calculated by the Gauss formula, is constrained by a minimum value $V_{\min}$. Then the whole optimization problem can be formulated as
\begin{equation}\label{eqn:formula_opt}
\left.
\begin{aligned}
&\mathop {\min }\limits_{\Gamma _{\rm w}}  \quad J = \int_{{\Gamma _{\rm{w}}}} {\int_\Xi  {m_J\bm v \cdot \bm n{f_1}} d\Xi d\Gamma } , \\
&{\rm{s}}{\rm{.t}}{\rm{.}}\quad Q = \frac{1}{D}\int_{{\Gamma _{\rm{w}}}} {\bm x \cdot \bm nd\Gamma }  + {V_{\min }} \le 0,
\end{aligned}
\right\}
\end{equation}
where $D$ is the spacial dimension of the problem (here we have $D=2$), and  $m_J$ is the moment variable for the objective we concern:
\begin{equation}
\begin{aligned}
& {m_J} = {v_1}\quad {\rm for~the~}{x_1}{\rm{ - direction~force~on~the~solid~body}},\\
& {m_J} = \frac{1}{2}{{\bm v}^2}\quad {\rm for~the~heat~on~the~solid~body}.
\end{aligned}
\end{equation}

The optimization problem \eqref{eqn:formula_opt} is written in the nested form, where the distribution function $\bm f$ and the solid boundary $\Gamma _{\rm w}$ satisfy the following gas-kinetic flow problem:
\begin{equation}\label{eqn:formula_bgk}
\left.
\begin{aligned}
\bm v \cdot \nabla \bm f - \frac{{{\bm g} - \bm f}}{{{\tau}}} &= 0 \quad {\rm{in}}\quad \Omega  \times \Xi, \\
\bm f - {\bm f_{\rm d}} &= 0\quad  {\rm{in}}\quad {\Gamma _{\rm d}} \times {\Xi ^ - },\\
\bm f - {\bm g_{\rm{w}}} &= 0\quad  {\rm{in}}\quad {\Gamma _{\rm w}} \times {\Xi ^ - },
\end{aligned}
\right\}
\end{equation}
where the boundary conditions for the far-field boundary $\Gamma _{\rm d}$ and the solid wall $\Gamma _{\rm w}$ are given in Section~\ref{sec:boundary}.

\subsection{Sensitivity analysis}\label{sec:sens}

The optimization problem  \eqref{eqn:formula_opt} is tackled using a gradient-based method. Consequently, we need to compute the derivative, or sensitivity, of the objective functional $J$ with respect to the design variables (i.e., the geometric parameters of $\Gamma _{\rm w}$). This is achieved through the adjoint analysis method.
Generally, there are two fashions of adjoint methods \cite{nadarajah2000comparison,peter2010numerical}. The first is the continuous adjoint approach, which involves deriving the adjoint equation and then discretizing it. The second is the discrete adjoint approach, which discretizes the primal equation first and then derives the discrete adjoint equation from the discretized primal equation. Here we initiate the development of our method following the continuous adjoint approach, although we will ultimately calculate the sensitivity using a combination of both.


According to Eq.~\eqref{eqn:formula_opt}, $J$ is determined by the distribution function $\bm f$ and the solid boundary $\Gamma _{\rm w}$, meanwhile $\bm f$ and $\Gamma _{\rm w}$ should satisfy the kinetic problem \eqref{eqn:formula_bgk}. Therefore, we introduce a set of Lagrangian multipliers (or in other words, adjoint variables) $\bm \phi ,{\bm \varphi _{\rm{d}}},{\bm \varphi _{\rm{w}}}$ to write the Lagrangian as
\begin{equation}\label{eqn:lagrangian}
{\cal L}(\bm f,\theta,\bm \phi,\bm \varphi _{\rm{d}},\bm \varphi _{\rm{w}} ) =  J + I + B_{\rm w} + B_{\rm d},
\end{equation}
where
\begin{equation}
I=\int_\Omega  {\int_\Xi  {\bm \phi  \cdot \left( {\bm v \cdot \nabla \bm f - \frac{{{\bm g} - \bm f}}{{{\tau}}}} \right)d\Xi } d\Omega },
\end{equation}
\begin{equation}
B_{\rm w}=\int_{{\Gamma _{\rm w}}} {\int_{{\Xi ^ - }} {{\bm \varphi _{\rm w}} \cdot \left( {\bm f - {\bm g_{\rm{w}}}} \right)d\Xi } d\Gamma },
\end{equation}
\begin{equation}
B_{\rm d}=\int_{{\Gamma _{\rm d}}} {\int_{{\Xi ^ - }} {{\bm \varphi _{\rm d}} \cdot \left( {\bm f - {\bm f_{\rm d}}} \right)d\Xi } d\Gamma }.
\end{equation}
The main obstacle to calculate the sensitivity with respect to the shape of $\Gamma _{\rm w}$ is that the change $\delta \Gamma _{\rm w}$ will also cause the change $\delta \bm f$ which has a contribution to the change of $J$. Fortunately, with the Lagrangian~\eqref{eqn:lagrangian} we can eliminate the contribution from $\delta \bm f$ by finding a set of adjoint variables $\bm \phi ,{\bm \varphi _{\rm{d}}},{\bm \varphi _{\rm{w}}}$ fulfilling
$d{\cal L}(\bm f;\delta \bm f) = 0$.
Then the total derivative of $J$ with respect to the shape of $\Gamma _{\rm w}$ can be calculated from the explicit expression of the partial derivative $d{\cal L}(\Gamma _{\rm w} ;\delta \Gamma _{\rm w} )$ since it is obvious that $\forall \bm \phi ,{\bm \varphi _{\rm{d}}},{\bm \varphi _{\rm{w}}}$ there is ${\cal L} \equiv J$.
Thus, the adjoint equation to determine the adjoint variables $\bm \phi ,{\bm \varphi _{\rm{d}}},{\bm \varphi _{\rm{w}}}$ can be established. The derivation involves the variational method and some mathematical arrangements. Finally the adjoint governing equation along with the boundary conditions can be formulated as
\begin{equation}\label{eqn:formula_adjointbgk}
\left.
\begin{aligned}
 - \bm v \cdot \nabla \bm \phi  = \frac{{{\bm \phi _{{\rm{eq}}}} - \bm \phi }}{{{\tau }}} + {\bm \phi _\tau }, \quad {\rm{in}}\quad \Omega  \times \Xi, \\
\bm \phi  = \bm 0, \quad {\rm{in}} \quad {\Gamma _{\rm d}} \times {\Xi ^ + },\\
\bm \varphi _{\rm{w}}  = \left( \begin{gathered}
  - \bm v \cdot \bm n\left( {{\phi _1} + {m_J}} \right)  \\
  - \bm v \cdot \bm n{\phi _2}
\end{gathered} \right),\quad {\rm{in}}\quad {\Gamma _{\rm w}} \times {\Xi ^ - },\\
\bm \phi  = \left( \begin{gathered}
 \int_{{\Xi ^ - }} \bm\varphi _{\rm{w}}  \cdot \frac{{\partial {{\bm g_{\rm{w}}}}}}{{\partial {F_{\rho{\rm{w}}}}}}d\Xi - m_J \\
 0
\end{gathered} \right),\quad {\rm{in}}\quad {\Gamma _{\rm w}} \times {\Xi ^ + },
\end{aligned}
\right\}
\end{equation}
where the collision-related terms ${\bm \phi _{{\rm{eq}}}},{\bm \phi _\tau }$ are
\begin{equation}\label{eqn:adjoint_eqdefines}
{\bm \phi _{\rm eq}} = \hat {\bm W} \cdot {\bf{\Psi }},\quad {\bm \phi _\tau } =  - {{\hat \rho }_\tau }\frac{1}{{{\tau }}}\frac{{\partial {\tau }}}{{\partial \bm W}} \cdot {\bf{\Psi }},
\end{equation}
with
\begin{equation}\label{eqn:adjoint_macdefines}
\hat {\bm W} = \int_\Xi  {\bm \phi  \cdot \frac{{\partial {\bm g }}}{{\partial {\bm W }}}d\Xi } , \quad {\hat \rho _\tau } = \int_\Xi  {\bm \phi  \cdot \frac{{{\bm g } - \bm f}}{{{\tau }}}d\Xi }. 
\end{equation}
Note that here we have canceled out the variable $\bm \varphi _{\rm{d}}$ since it is not used in the subsequent calculations. It can be seen that the moments factor to get the adjoint macroscopic variable $\hat {\bm W}$ is $\partial {\bm g }/\partial {\bm W }$, by analogy with the moments tensor $\bf{\Psi }$ for the primal gas-kinetic governing equation. It is also worth noting that in the adjoint problem \eqref{eqn:formula_adjointbgk}, $m_J$ plays the role as a constant flux source on the gas-solid interface, reflecting the influence of the objective functional.

On obtaining the adjoint variable $\bm \varphi _{\rm{w}}$ from the adjoint problem \eqref{eqn:formula_adjointbgk}, directly taking the partial differential of the Lagrangian \eqref{eqn:lagrangian} about the variation $\delta \Gamma _{\rm w}$ will yield
\begin{equation}\label{eqn:sens_shape}
d{\cal L}(\Gamma _{\rm w} ;\delta \Gamma _{\rm w} ) = dJ(\Gamma _{\rm w} ;\delta \Gamma _{\rm w} )+dB_{\rm w}(\Gamma _{\rm w};\delta \Gamma _{\rm w} ), 
\end{equation}
which is just the total differential of the objective $J$ about the variation $\delta \Gamma _{\rm w}$ and the corresponding sensitivity can be determined. Note that the partial differentials of $I$ and $B_{\rm d}$ about the variation $\delta \Gamma _{\rm w}$ are just 0 since the flow variable $\bm f$ is unchanged.

We will discretize separately the primal gas-kinetic equation \eqref{eqn:formula_bgk} and the adjoint equation \eqref{eqn:formula_adjointbgk} using multiscale numerical schemes that are accurate for both rarefied and continuum flows. The choice to adopt a manner of continuous adjoint  is primarily driven by the ease of implementation, as has been discussed  in our previous work \cite{yuan2024design}.
Once we have determined the adjoint variable $\bm \varphi _{\rm{w}}$, to calculate the sensitivity with respect to the geometry parameters of $\Gamma _{\rm w}$, we will analyze Eq. \eqref{eqn:sens_shape} on the solid boundary $\Gamma _{\rm w}$
using a method akin to the discrete adjoint approach. Details will be provided in the subsequent sections.

\section{Numerical method}\label{sec:method}

\subsection{Numerical discretization for the primal and adjoint equations}\label{sec:method_scheme}

The primal equation~\eqref{eqn:formula_bgk} and the adjoint equation~\eqref{eqn:formula_adjointbgk} are discretized by the finite-volume discrete velocity method with second-order accuracy in physical space:
\begin{equation}\label{eqn:numerpr_gov}
\sum\limits_{j \in N\left( i \right)} {{A_{ij}}{\bm v_k} \cdot {\bm n_{ij}}\bm f_{ij,k}}  = {V_i}\frac{{\bm g_{i,k} - \bm f_{i,k}}}{{\tau _{i}}},
\end{equation}
\begin{equation}\label{eqn:numerad_gov}
 - \sum\limits_{j \in N\left( i \right)} {{A_{ij}}{\bm v_k} \cdot {\bm n_{ij}}\bm \phi _{ij,k}}  =  {V_i}\frac{{\bm \phi _{{\rm eq},i,k} - \bm \phi _{i,k}}}{{\tau _{i}}} + {V_i}\bm \phi _{\tau ,i,k} ,
\end{equation}
where the super/subscripts $i,k$ correspond to the discretizations in physical
space and velocity space respectively; $j$ denotes the neighboring cell of cell $i$ and $N\left( i \right)$ is the set of all of the neighbors of $i$; $ij$ denotes the variable at the interface between cell $i$ and $j$; $A_{ij}$ is the interface area, ${\bm n_{ij}}$ is the outward normal unit vector of interface $ij$ relative to cell $i$, and ${V_i}$ is the control volume of cell $i$.

In order to solve the discretized governing equations accurately and efficiently for both rarefied and continuum gas flows, the implicit multiscale gas-kinetic schemes with the prediction acceleration technique is applied, which have the following features:
\begin{enumerate}
\item The interface distribution function $\bm f _{ij,k}$ and $\bm \phi _{ij,k}$ are treated by the idea of the discrete unified gas-kinetic scheme \cite{guo2013discrete} to guarantee the accuracy of the scheme in different flow regimes.
\item Approximate macroscopic equations constructed from the continuum limit are solved to provide prediction solutions to accelerate the convergence of the primal and adjoint gas-kinetic numerical systems, saving a lot of computational cost especially for the continuum flow calculation. 
\end{enumerate}
The detailed computation procedures of the schemes are hard to describe in a few words and are generally the same with those presented in \cite{yuan2024design}, therefore here we omit them for conciseness. We mention that although the numerical schemes are a bit complex, they are quite important for achieving good accuracy and efficiency in all flow regimes \cite{zhu2017performance,yuan2020conservative,yuan2021novel}. 
After solving Eqs.~\eqref{eqn:numerpr_gov} and \eqref{eqn:numerad_gov}, the gas flow variable $\bm f$ and the adjoint variable $\bm \phi,\bm \varphi _{\rm{w}}$ are determined.

\subsection{Sensitivity with respect to the boundary mesh nodes}\label{sec:sens_nod}

After obtaining the flow variable $\bm f$ and the adjoint variable $\bm \varphi _{\rm{w}}$, we use Eq.~\eqref{eqn:sens_shape} to calculate the sensitivity of the Lagrangian $\cal L$, which is also the sensitivity of the objective $J$, with respect to the nodes' coordinates determining the shape of the solid boundary $\Gamma _{\rm w}$.

The analysis about Eq.~\eqref{eqn:sens_shape} takes a method  similar to the discrete adjoint analysis. First, for the two terms $J$ and $B_{\rm w}$ involved in Eq.~\eqref{eqn:sens_shape}, their discretized forms can be written as
\begin{equation}\label{eqn:numer_J}
\left.
\begin{aligned}
J = \sum\limits_k {\sum\limits_l {{J_{l,k}}\Delta {\Xi _k}} } ,\\
{J_{l,k}} = {m_{J,k}}{{\bm v}_k} \cdot {{\bm n}_l}{f_{1,l,k}}{A_l},
\end{aligned}
\right\}
\end{equation}
and 
\begin{equation}\label{eqn:numer_B}
\left.
\begin{aligned}
{B_{\rm{w}}} = \sum\limits_{{{\vec v}_k} \in {\Xi ^ - }} {\sum\limits_l {{B_{{\rm{w}},l,k}}\Delta {\Xi _k}} },\\
{B_{{\rm{w}},l,k}} = {\bm \varphi _{{\rm{w}},l,k}} \cdot \left( {{{\bm f}_{l,k}} - {{\bm g}_{{\rm w},l,k}}} \right){A_l}.
\end{aligned}
\right\}
\end{equation}
Note that according to the discretization for the governing equations \eqref{eqn:numerpr_gov} and \eqref{eqn:numerad_gov}, $\Gamma _{\rm w}$ has been discretized into several surface elements (segments for 2D) denoted by $l$. The area, centroid and outward normal unit vector (pointing to the solid side) of the $l$-th element are denoted by $A_l,\bm x_l$ and $\bm n_l$, respectively. The variables ${\bm f}_{l,k},{\bm g}_{{\rm w},l,k}$ and $\bm \varphi _{{\rm{w}},l,k}$ are defined at the center of the surface element, and their values should have been all determined by solving Eqs.~\eqref{eqn:numerpr_gov} and \eqref{eqn:numerad_gov}.
It is particularly important to note that, here ${\bm g}_{{\rm w},l,k}$ should be calculated by the no-mass-penetration condition in the discretization level, i.e.~the discretized version for Eq.~\eqref{eqn:formula_bgk_fdw}, and it can be written as
\begin{equation}
{{\bm g}_{{\rm{w}},l,k}} = {F_{\rho {\rm{w}},l}}{{\bar{\bm g}}_{{\rm{w}},l,k}},
\end{equation}
where ${F_{\rho {\rm{w}},l}}$ is calculated by the numerical integration in the velocity space
\begin{equation}
{F_{\rho {\rm{w}},l}} = \sum\limits_{{{\bm v}_k} \in {\Xi ^ + }} {{{\bm v}_k} \cdot {{\bm n}_l}{f_{1,l,k}}\Delta {\Xi _k}},
\end{equation}
and ${{\bar {\bm g}}_{{\rm{w}},l,k}}$ is the normalization
\begin{equation}\label{eqn:barg}
{{\bar {\bm g}}_{{\rm{w}},l,k}} = \frac{{{{\bm g}_{\rm{M}}}(1,\bm 0,{T_{\rm{w}}})}}{{ - \sum\limits_{{{\bm v}_k} \in {\Xi ^ - }} {{{\bm v}_k} \cdot {{\bm n}_l}{g_{{\rm{M}},1}}(1,\bm 0,{T_{\rm{w}}})\Delta {\Xi _k}} }}.
\end{equation}

Investigating the discretized forms of $J$ and $B_{\rm w}$, and considering that we don't expect the topological change of the spatial discretization, we find that the impact of the variation $\delta {\Gamma _{\rm w}}$ on the $l$-th boundary surface element can be divided into three parts: the area change $dA_l$, the centroid location change $d\bm x_l$ and the change of the outward normal unit vector $d\bm n_l$. Calculating the sensitivity about the variation $\delta {\Gamma _{\rm w}}$ is then turned into calculating the derivatives of $J_{l,k},B_{{\rm{w}},l,k}$ with respect to $A_l,\bm x_l,\bm n_l$. 

For $J_{l,k}$ which has the expression \eqref{eqn:numer_J}, its derivatives can be derived as
\begin{equation}\label{eqn:sens_J_itf}
\left.
\begin{aligned}
\frac{{\partial {J_{l,k}}}}{{\partial A_l}} &= {m_{J,k}}\bm v_k \cdot \bm n_l{f_{1,l,k}},\\
\frac{{\partial {J_{l,k}}}}{{\partial {{\bm x}_l}}} &= {m_{J,k}}\bm v_k \cdot \bm n_l{A_l}\frac{{\partial {f_{1,l,k}}}}{{\partial {{\bm x}_l}}},\\
\frac{{\partial {J_{l,k}}}}{{\partial {{\bm n}_l}}} &= {m_{J,k}}\bm v_k{A_l}{f_{1,l,k}},
\end{aligned}
\right\}
\end{equation}
where the spatial derivative ${\partial {f_{1,l,k}}}/{{\partial {{\bm x}_l}}}$ will be calculated by the reconstruction through Gauss formula: suppose the cell $i$ is adjacent to the surface element $l$, then
\begin{equation}\label{eqn:gaussgrad}
\frac{{\partial {f_{1,l,k}}}}{{\partial {{\bm x}_l}}} = \frac{1}{{{V_i}}}\sum\limits_{j \in N(i)} {{f_{1,ij,k}}{{\bm n}_{ij}}{A_{ij}}} ,
\end{equation}
where $f_{1,ij,k}$ at the interface should have already been determined in solving the primal equation~\eqref{eqn:numerpr_gov}.

The derivation of the derivatives for $B_{{\rm{w}},l,k}$ necessitates a slightly more sophisticated technical approach.
We mention the following two points:
\begin{enumerate}
\item The integration ranges ${\Xi ^ + },{\Xi ^ - }$ in the velocity space will change with $d\bm n_l$, but fortunately the corresponding integrands all have the factor $\bm v \cdot \bm n$, which is $0$ at the bounds of the integral. Therefore the change of the integration range can be simply ignored.
\item The normalized Maxwellian ${{\bar {\bm g}}_{{\rm{w}},l,k}}$ \eqref{eqn:barg} will change with $d\bm n_l$, and the corresponding derivative can be derived as
\begin{equation}
\frac{{\partial {{\bar {\bm g} }_{{\rm{w}},l,k}}}}{{\partial {{\bm n}_l}}} = {\bar {\bm g} _{{\rm{w}},l,k}}\sum\limits_{{{\bm v}_k} \in {\Xi ^ - }} {{{\bm v}_k}{{\bar g }_{{\rm{w,1}},l,k}}{\Delta {\Xi _k}}} .
\end{equation}
\end{enumerate}
After some mathematical deduction, the derivatives for $B_{{\rm{w}},l,k}$ of Eq.~\eqref{eqn:numer_B} can be arranged as
\begin{equation}\label{eqn:sens_B_itf}
\left.
\begin{aligned}
\frac{{\partial {B_{{\rm{w}},l,k}}}}{{\partial {A_l}}} &= 0,\\
\frac{{\partial {B_{{\rm{w}},l,k}}}}{{\partial {{\bm x}_l}}} &= {{\bm \varphi }_{{\rm{w}},l,k}} \cdot \frac{{\partial {{\bm f}_{l,k}}}}{{\partial {{\bm x}_l}}}{A_l} - {{\bm \varphi }_{{\rm{w}},l,k}} \cdot {{\bar {\bm g}}_{{\rm{w}},l,k}}{A_l}\left( {\sum\limits_{{{\bm v}_k} \in {\Xi ^ + }} {{{\bm v}_k} \cdot {{\bm n}_l}\frac{{\partial {f_{1,l,k}}}}{{\partial {{\bm x}_l}}}\Delta {\Xi _k}} } \right),\\
\frac{{\partial {B_{{\rm{w}},l,k}}}}{{\partial {{\bm n}_l}}} &=  - {{\bm \varphi }_{{\rm{w}},l,k}} \cdot {{\bar {\bm g}}_{{\rm{w}},l,k}}\left( {\sum\limits_{k } {{{\bm v}_k}{f_{1,l,k}}\Delta {\Xi _k}} } \right),
\end{aligned}
\right\}
\end{equation}
where the spatial derivative ${\partial {\bm f_{l,k}}}/{{\partial {{\bm x}_l}}}$ is calculated by the method similar to
Eq.~\eqref{eqn:gaussgrad}.

Based on Eqs.~\eqref{eqn:sens_J_itf} and \eqref{eqn:sens_B_itf}, the derivatives for $J$ and $B_{\rm w}$ can be simply calculated by summation according to Eqs.~\eqref{eqn:numer_J} and \eqref{eqn:numer_B}. Finally the sensitivity for the Lagrangian $\cal L$ with respect to the surface elements' attributes $A_l,\bm x_l,\bm n_l$ can be obtained according to Eq.~\eqref{eqn:sens_shape}. For the sake of generality, this sensitivity can be further transformed into that with respect to the coordinates of the nodes defining the boundary ${\Gamma _{\rm w}}$. The transformation only involves the differentiation of geometric relationships. In 2D case, suppose the $l$-th segment has 2 nodes with the coordinates $\bm x_{l,1},\bm x_{l,2}$, the transformation derivatives can be formulated as
\begin{equation}\label{eqn:sens_nod}
\left.
\begin{aligned}
\frac{{\partial {A_l}}}{{\partial {{\bm x}_{l,1}}}} &= \frac{{{{\bm x}_{l,1}} - {{\bm x}_{l,2}}}}{{\left| {{{\bm x}_{l,1}} - {{\bm x}_{l,2}}} \right|}},\\
\frac{{\partial {{\bm x}_l}}}{{\partial {{\bm x}_{l,1}}}} &= \frac{1}{2},\\
\frac{{\partial {{\bm n}_l}}}{{\partial {{\bm x}_{l,1}}}} &= {\rm sign}(({{\bm x}_{l,1}} - {{\bm x}_{l,2}}) \times {{\bm n}_l})\frac{1}{{{A_l}}}\left( {\begin{array}{*{20}{c}}
{ - {n_{l,1}}{n_{l,2}}}&{ - n_{l,2}^2}\\
{n_{l,1}^2}&{{n_{l,1}}{n_{l,2}}}
\end{array}} \right).
\end{aligned}
\right\}
\end{equation}
Then the sensitivity of $\cal L$ (also $J$) with respect to the coordinates of the nodes defining ${\Gamma _{\rm w}}$ can be obtained.

It is worth noting that for the gradient-based optimizer, in addition to the sensitivity of the objective, it is also required to calculate the sensitivity of the constraint. If the constraint is dependent on the flow variable $\bm f$ then the corresponding adjoint analysis for its sensitivity should also be performed. But in the present work we only consider the volume constraint $Q$ in Eq.~\eqref{eqn:formula_opt}, whose sensitivity is only dependent on the geometry of the boundary ${\Gamma _{\rm w}}$ and can be explicitly derived as
\begin{equation}\label{eqn:sens_con_nod}
\frac{{\partial Q}}{{\partial {{\bm x}_m}}} = \sum\limits_{z \in {N_{\rm{s}}}(m)} {\frac{1}{D}{A_{m,z}}{{\bm n}_{m,z}}},
\end{equation}
where ${\bm x}_m$ is the coordinate of the $m$-th mesh node defining ${\Gamma _{\rm w}}$; ${N_{\rm{s}}}(m)$ is the set of ${\Gamma _{\rm w}}$'s surface elements which contain the node $m$; ${A_{m,z}},{\bm n}_{m,z}$ is the area and outward normal unit vector (pointing to the solid side) of the surface element; $D$ is the dimension of the physical space.

\subsection{The CST parameterization}\label{sec:cst}
Directly optimizing the location of the nodes defining ${\Gamma _{\rm w}}$ may lead to irregular or physically unrealistic shape of the solid, such as the wavelike or stair-step pattern. It is common to restrict the design variables to a design space of smooth, regular shapes by some parameterization process. Later in Section \ref{sec:case0} we will see this parameterization treatment is particularly necessary for the optimization adopting a rarefied gas flow solver of the discrete velocity method.

We adopt the parameterization method of the class function/shape function transformation (CST) \cite{kulfan2008universal,bu2013aerodynamic} for the shape optimization of airfoils. The surface of the airfoil is defined as
\begin{equation}\label{eqn:cstsuf}
\left.
\begin{aligned}
{\rm{upper~surface}}:\quad {\eta _{\rm{u}}} &= C(\xi ){S_{\rm{u}}}(\xi ) + \xi {\eta _{{\rm{T,u}}}},\\
{\rm{lower~surface}}:\quad {\eta _{\rm{l}}} &= C(\xi ){S_{\rm{l}}}(\xi ) + \xi {\eta _{{\rm{T,l}}}},
\end{aligned}
\right\}
\end{equation}
where the subscripts u,l denote the upper or lower surface; $\xi,\eta$ are the horizontal/vertical coordinates of the airfoil geometry normalized by the chord length; $C(\xi )$ is the so-called class function defined as
\begin{equation}\label{eqn:cst_classfunc}
C(\xi ) = {\xi ^{N1}}{\left( {1 - \xi } \right)^{N2}};
\end{equation}
$S_{\rm{u}}(\xi),S_{\rm{l}}(\xi)$ are the shape functions defined as
\begin{equation}\label{eqn:cst_shapefunc}
\left.
\begin{aligned}
{S_{\rm{u}}}(\xi ) = \sum\limits_{\beta = 0}^{N_{\rm B}} {{A_{{\rm{u}},\beta}}{S_\beta}(\xi )} ,\\
{S_{\rm{l}}}(\xi ) = \sum\limits_{\beta = 0}^{N_{\rm B}} {{A_{{\rm{l}},\beta}}{S_\beta}(\xi )} ,
\end{aligned}
\right\}
\end{equation}
where $N_{\rm B}$ is the prescribed order of the shape function polynomial and ${{S_\beta}(\xi )}$ is the Bernstein polynomial:
\begin{equation}
{S_\beta}(\xi ) = \frac{{{N_{\rm B}}!}}{{\beta!({N_{\rm B}} - \beta)!}}{\xi ^\beta}{\left( {1 - \xi } \right)^{{N_{\rm B}} - \beta}}.
\end{equation}
In sum, the design variables for the above parameterization are: $N1,N2$ defining the general class of the airfoil geometry, $A_{{\rm{u}},\beta},A_{{\rm{l}},\beta}$ defining the specific shape of the upper/lower surface, and $\eta _{{\rm{T,u}}},\eta _{{\rm{T,l}}}$ controlling the trailing-edge thickness. For ${N_{\rm B}}$-th order of the Bernstein polynomial, the number of the design variables will be $2{N_{\rm B}}+6$. In the current study we choose ${N_{\rm B}}=6$, which is a common setup in the airfoil optimization.

At the beginning of optimization, the design variables for the initial shape should be determined. This can be finished by solving the following least-squares problem:
\begin{equation}\label{eqn:cstini}
\min \quad \sum\limits_m {{{\left( {{{\bm x}_{{\rm{CST}}}}({\xi _m}) - {{\bm x}_{{\rm{Initial}},m}}} \right)}^2}} ,
\end{equation}
where ${{{\bm x}_{{\rm{Initial}},m}}}$ is the given $m$-th initial mesh node defining ${\Gamma _{\rm w}}$, which has a normalized horizontal coordinate of ${{\xi _m}}$, and ${{{\bm x}_{{\rm{CST}}}}({\xi _m})}$ is the corresponding coordinate calculated by the CST method at ${{\xi _m}}$.
We assume the normalized horizontal coordinate ${{\xi _m}}$ for the initial node will stay fixed throughout the optimization, and when the design variables have an update only the normalized vertical coordinate ${{\eta _m}}$ is changed.
After we obtain the sensitivity with respect to the coordinates of the nodes defining ${\Gamma _{\rm w}}$ by the method in Section \ref{sec:sens_nod}, we can further transform it to the sensitivity with respect to the CST design variables by the chain rule, which only involves some simple explicit expressions and is omitted here for conciseness.

\subsection{Overall algorithm framework of the optimization}\label{sec:method_optframework}

The present optimization method follows the general procedure of the gradient-based optimization method.
The whole optimization procedure is listed as follows:
\begin{description}

   \item[Step 1.] Initialize the CST design variables by solving the least-squares problem \eqref{eqn:cstini}.
    
	\item[Step 2.] Use the initialized/updated design variables to obtain the updated mesh nodes of the solid boundary $\Gamma_{\rm w}$. Then the mesh deformation in the bulk region is completed by the radial-basis-function interpolation method \cite{rendall2009efficient}. We choose the Wendland's C0 function as the basis function since it performs better than the popular Wendland's C2 function in the case of the thin airfoil deformation concerned by us.

    \item[Step 3.] Solve the primal gas-kinetic equation \eqref{eqn:numerpr_gov} to obtain the primal flow variable $\bm f$, and then solve the adjoint equation \eqref{eqn:numerad_gov} to get the adjoint variable ${\bm \varphi _{\rm{w}}}$. The value of the objective $J$ and the constraint $Q$ can be meanwhile determined.

    \item[Step 4.] Calculate the sensitivities of the objective $J$ and the constraint $Q$ by the method described in Section \ref{sec:sens_nod} and Section \ref{sec:cst}.
    
    \item[Step 5.] Substitute the values of the objective $J$ and the constraint $Q$, and their corresponding sensitivities into the gradient-based optimizer and get the updated design variables. The optimizer here we employ is the sequential least-squares quadratic programming (SLSQP) algorithm implemented by Johnson in his NLopt library \cite{johnson2007nLopt}. This is a SQP algorithm based on the quasi-Newton method with the BFGS formula, where the constraint is handled by a primal/dual method, as described in Refs.~\cite{kraft1988software,lawson1995solving}.
    
    \item[Step 6.] Judge whether the optimization converges by the criterion
\begin{equation}\label{eqn:criterion_opt}
\frac{{\left| {{J^{\rm new}} - {J^{\rm old}}} \right|}}{{\left| {{J^{\rm old}}} \right|}} < 3 \times {10^{ - 5}},
\end{equation}
where ${J^{\rm old}}$ and ${J^{\rm new}}$ are the values of the objective before and after an optimization step. If the criterion is not met then go to Step 2 to start a new step of optimization. It is found in our current work that the criterion \eqref{eqn:criterion_opt} is strict enough to find the optimum.

\end{description}

\begin{figure}[h]
	\centering
	\includegraphics[width=0.95\textwidth]{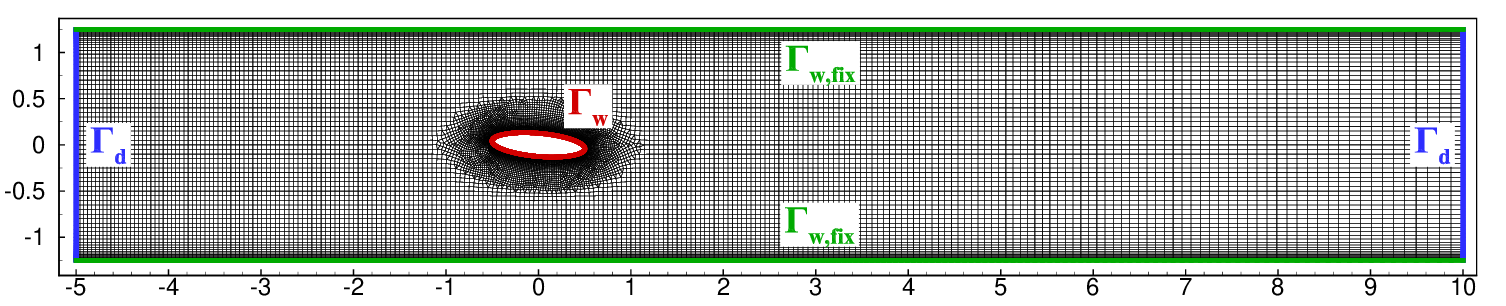}
	\caption{\label{fig:case0_mesh}The setups of mesh and boundary condition for the flow passing over an elliptic cylinder inside a channel, 29384 cells in total.}
\end{figure}

\section{Numerical results}\label{sec:test}

\subsection{Validation of the sensitivity}\label{sec:case0}

The accuracy of the sensitivity calculated by our method will be verified.
The flow passing over an elliptic cylinder inside a channel is simulated, and the sensitivity of the drag exerted on the cylinder is calculated for validation, namely the objective function is $J$ in Eq.~\eqref{eqn:formula_opt} with $m_{J}=v_1$. The mesh and boundary condition are illustrated in Fig.~\ref{fig:case0_mesh}. The channel has a domain of $\{ \bm x|{x_1} \in [ - 5,10],{x_2} \in [ - 1.25,1.25]\} $ where the center of the cylinder is placed at $(0,0)^\top$. The elliptic cylinder has a unit chord length of $c=1$ and a thickness of $0.25$, and the angle of attack is $5^\circ$. At the inlet/outlet boundary ${\Gamma _{\rm d}}$ the following Dirichlet boundary condition is imposed:
\begin{equation}\label{eqn:drlt_maxwell}
\bm f = {\bm g_{\rm M}}({\rho _\infty },{\bm u_\infty },{T_\infty })\quad {\rm{in}}\quad {\Gamma _{\rm d}} \times {\Xi ^ - },
\end{equation}
where the state variables ${\rho _\infty },{\bm u_\infty },{T_\infty }$ correspond to a Mach number of 0.6. At the upper/lower wall ${\Gamma _{\rm w,fix}}$ and the cylinder surface ${\Gamma _{\rm w}}$, the diffuse boundary condition as described by Eq.~\eqref{eqn:formula_bgk_fdw0} is applied, which has the condition ${\bm u_{\rm w}} = 0,{T_{\rm w}} = {T_\infty }$. Note that here ${\Gamma _{\rm w,fix}}$ is excluded from the design and the objective calculation, namely we only investigate the sensitivity of the drag on ${\Gamma _{\rm w}}$ with respect to the shape of ${\Gamma _{\rm w}}$.
The Knudsen numbers considered are ${\rm Kn}=0.001,0.1,10$,  which is defined by Eq.~\eqref{eqn:kndefine} with the inlet condition ${\rho _\infty },{T_\infty }$ and the reference length $c$; this covers the gas flows from continuum regime to free-molecular regime.
The physical computational domain is discretized by a nonuniform unstructured mesh with 29384 cells in total, where the mesh size near the upper/lower wall and the cylinder is refined. Specifically, the mesh height is $10^{-4}c$ near the cylinder to make sure the mesh independence is achieved.

\begin{figure}[t]
	\centering
	\subfigure[Sensitivity with respect to $x_1$]{\includegraphics[width=0.47\textwidth]{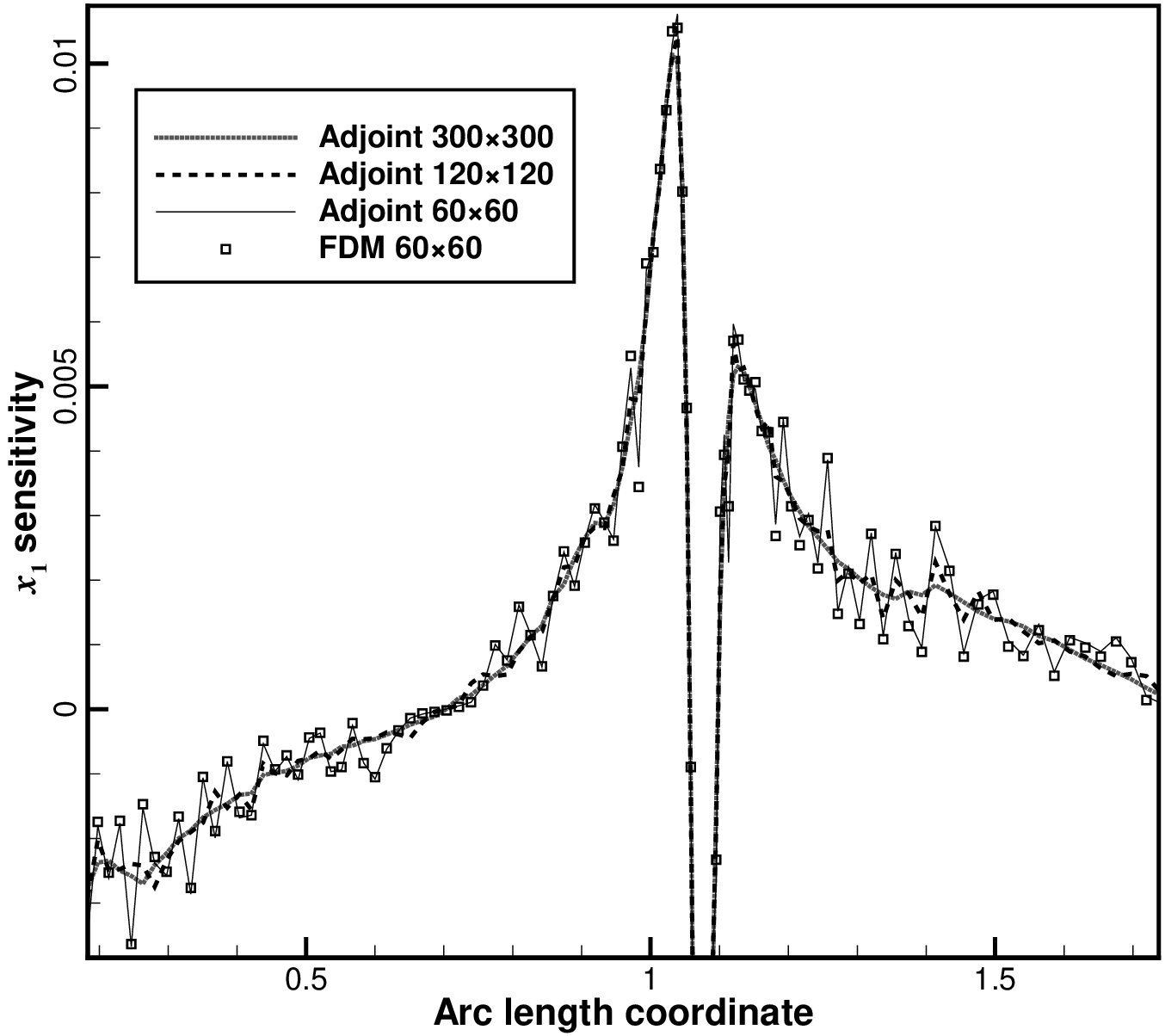}}\hspace{0.02\textwidth}
	\subfigure[Sensitivity with respect to $x_2$]{\includegraphics[width=0.47\textwidth]{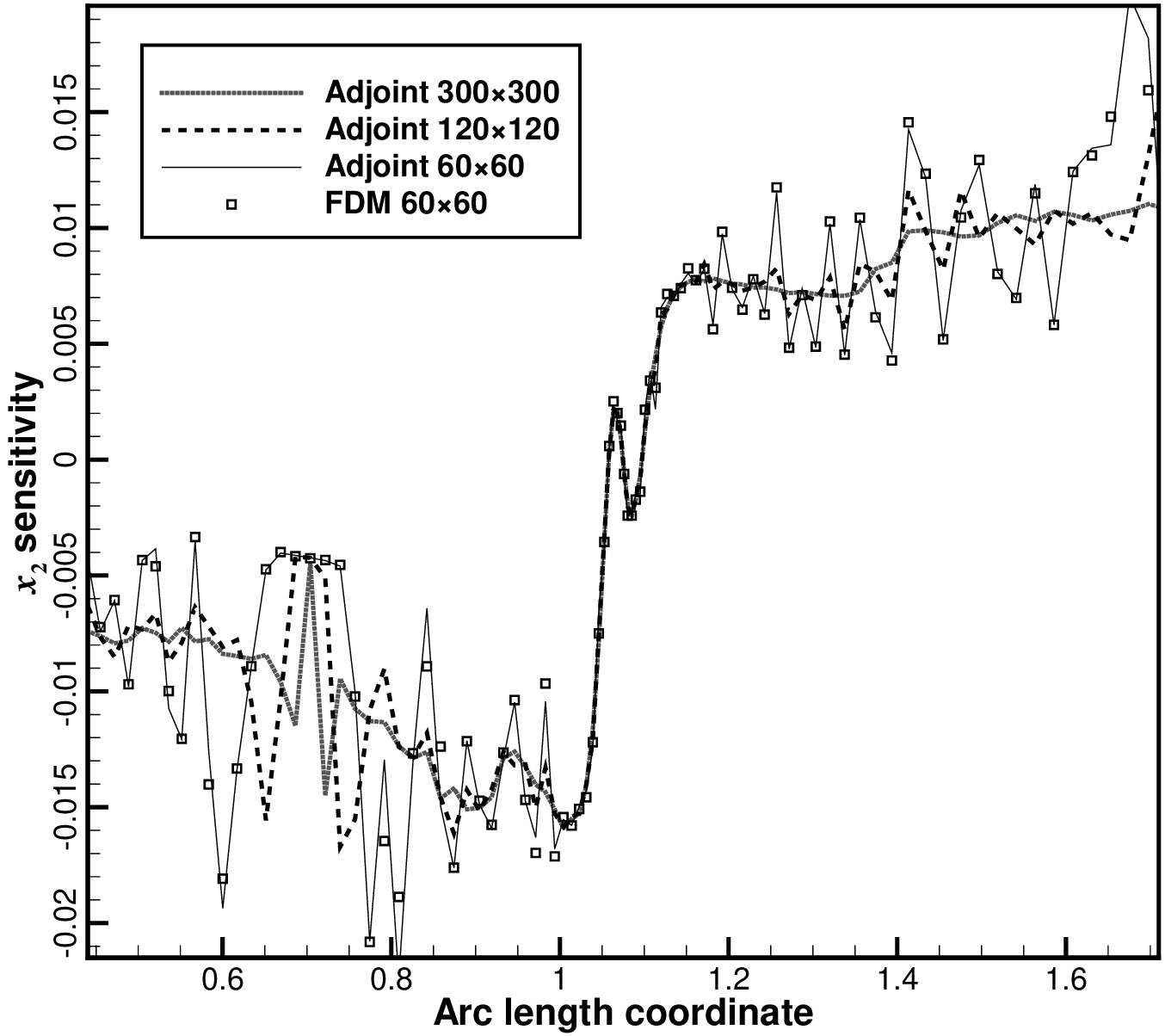}}
	\caption{\label{fig:case0_sensx} Flow over an elliptic cylinder inside a channel at ${\rm Kn}=10$: sensitivity of the drag with respect to the coordinates of the mesh points composing the cylinder surface ${\Gamma _{\rm w}}$. Computed by our method and FDM, with the different resolutions of uniform discretization for molecular velocity: $60 \times 60,120 \times 120$ and $300 \times 300$. The arc length is calculated counterclockwise from the leading edge point of the cylinder.}
\end{figure}

First we investigate the sensitivity with respect to the coordinates $\bm x_m=({x_{m,1}},{x_{m,2}})^\top$ of the mesh points composing the cylinder surface ${\Gamma _{\rm w}}$. We compare the sensitivity obtained by our method with that calculated by the finite difference method (FDM) through
\begin{equation}\label{eqn:fdm_x}
	\begin{aligned}
{\left( {\frac{{\partial J}}{{\partial {x_{m,1}}}}} \right)^{{\rm{FDM}}}} =& \frac{1}{2\epsilon}J( \ldots ,{{\bm x}_{m - 1}},{x_{m,1}} + \epsilon,{x_{m,2}},{{\bm x}_{m + 1}}, \ldots ) \\
-& \frac{1}{2\epsilon}J( \ldots ,{{\bm x}_{m - 1}},{x_{m,1}} - \epsilon,{x_{m,2}},{{\bm x}_{m + 1}}, \ldots ) ,
\end{aligned}
\end{equation}
where $\epsilon$ is the step size for the central difference. The results for ${\rm Kn}=10$ are shown in Fig.~\ref{fig:case0_sensx}. Here for the adjoint method, three different discretizations for the molecular velocity space are adopted, i.e., $60 \times 60,120 \times 120$ and $300 \times 300$ uniform meshes in the range $[ - 6{a_\infty },6{a_\infty }]$ where $a_\infty$ is the far-field acoustic speed. In FDM, due to the high computational cost, only the $60 \times 60$ velocity discretization is adopted. From Fig.~\ref{fig:case0_sensx} we can see:
\begin{enumerate}
\item The results of the adjoint method and FDM agree well when under the same velocity discretization, which verifies the accuracy of our method.
\item The profile of the sensitivity with respect to the coordinates of the points exhibit severe oscillation. This oscillation is gradually suppressed along with the refining of the velocity discretization, but can not be completely eliminated even under a very fine $300 \times 300$ velocity discretization.
\end{enumerate}

\begin{figure}
\centering
\includegraphics[width=0.47\textwidth]{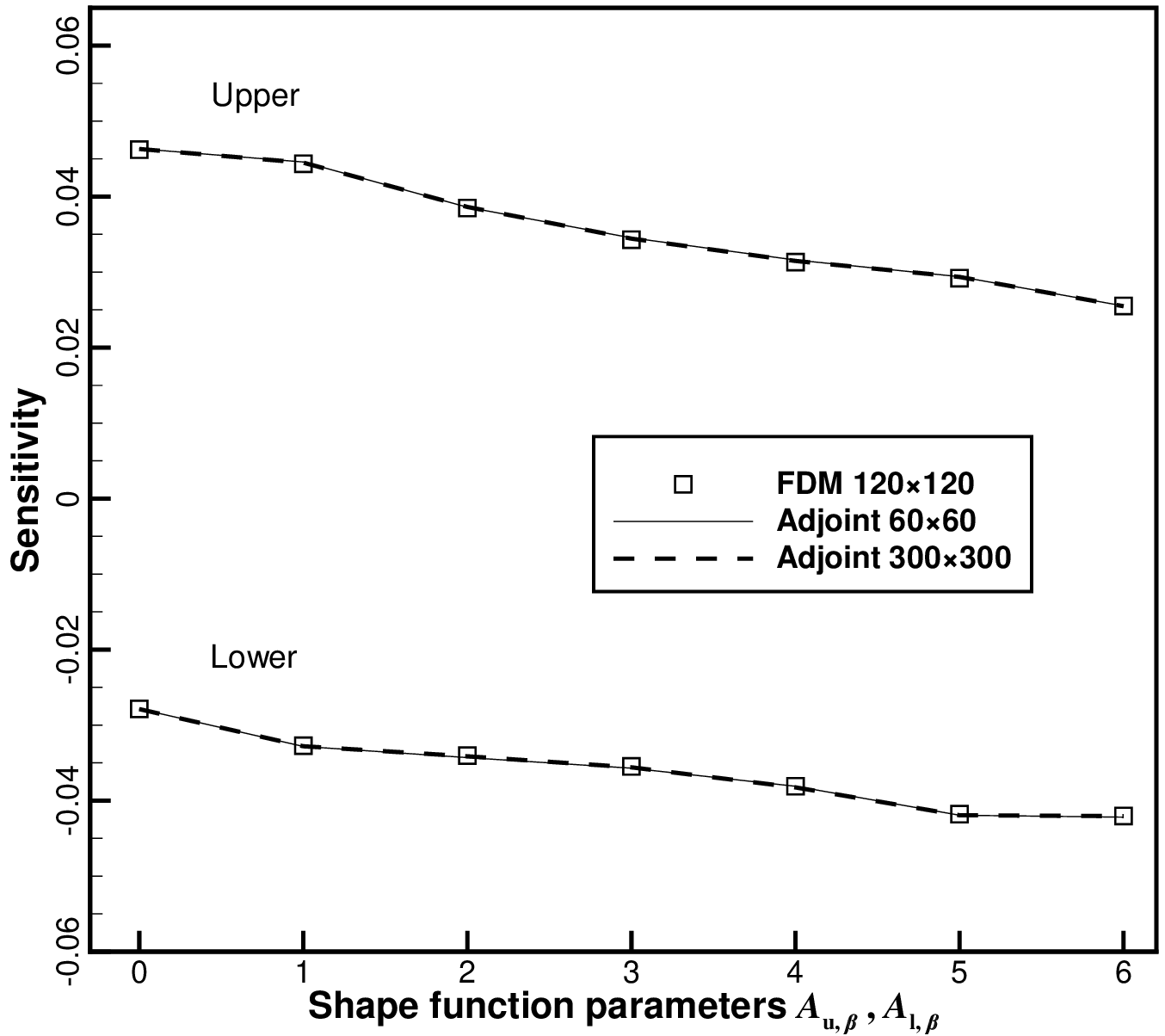}
\caption{\label{fig:case0_kn10_senscst} Flow over an elliptic cylinder inside a channel at ${\rm Kn}=10$: sensitivity of the drag with respect to the CST shape function parameters $A_{{\rm{u}},\beta},A_{{\rm{l}},\beta}$ defined in Eq.~\eqref{eqn:cst_shapefunc}. Computed by our method and FDM, with the different resolutions of uniform discretization for molecular velocity: $60 \times 60,120 \times 120$ and $300 \times 300$.}
\end{figure}

Clearly, the oscillation in the sensitivity profile is highly undesirable for optimization, as it can result in bumpy and irregular surfaces. Moreover, this oscillation appears to be related to the discretization of the velocity phase space. However, even when we refine the velocity discretization to a high resolution of $300 \times 300$, which is computationally very expensive for the discrete velocity method, the sensitivity profile does not fully converge and continues to exhibit oscillations.
Fortunately, in our practice we find that some parametric geometry representation approaches employed in the traditional shape optimization methods can effectively remove these oscillations. It seems that these oscillations are some white noises and the parameterization process works just like a low-pass filter, and the original design space populated with bumpy geometries is trimmed into a design space with smooth geometries. Specifically, in this paper we adopt the CST parameterization as described in Section \ref{sec:cst}. The results of the sensitivity with respect to the CST design variables are shown in Fig.~\ref{fig:case0_kn10_senscst} and Table \ref{tab:case0_kn10_senscst}. The results of FDM are calculated by perturbing the design variable and then doing the central difference~\eqref{eqn:fdm_x}. It is seen that the sensitivity with respect to the shape function parameters has a smooth profile, and the sensitivity obtained from the $60\times60$ velocity discretization has already reached the mesh independence. Meanwhile, the sensitivities computed by our adjoint method and FDM agree well with each other.

\begin{table}[t]
\centering
\caption{\label{tab:case0_kn10_senscst}Flow over an elliptic cylinder inside a channel at ${\rm Kn}=10$: sensitivity of the drag with respect to the CST design variables $N1,N2,\eta _{{\rm{T,u}}},\eta _{{\rm{T,l}}}$ defined in Eq.~\eqref{eqn:cstsuf}. Computed by our method and FDM, with the different resolutions of uniform discretization for molecular velocity.} \vspace{0.3cm}
\begin{tabular}{cccccc}
\hline
Method                   & \begin{tabular}[c]{@{}c@{}}Velocity\\ Discretization\end{tabular} & $N1$      & $N2$      & $\eta _{{\rm{T,u}}}$    & $\eta _{{\rm{T,l}}}$     \\ \hline
FDM                      & $120\times120$                                                               & -0.1254 & -0.1178 & 0.2829 & -0.3949 \\ \hline
\multirow{3}{*}{Adjoint} & $60\times60$                                                                & -0.1257 & -0.1182 & 0.2840 & -0.3962 \\
                         & $120\times120$                                                               & -0.1256 & -0.1180 & 0.2835 & -0.3957 \\
                         & $300\times300$                                                               & -0.1256 & -0.1181 & 0.2835 & -0.3958 \\ \hline
\end{tabular}
\end{table}

\begin{figure}[t]
\centering
\subfigure[${\rm Kn}=0.1$]{\includegraphics[width=0.47\textwidth]{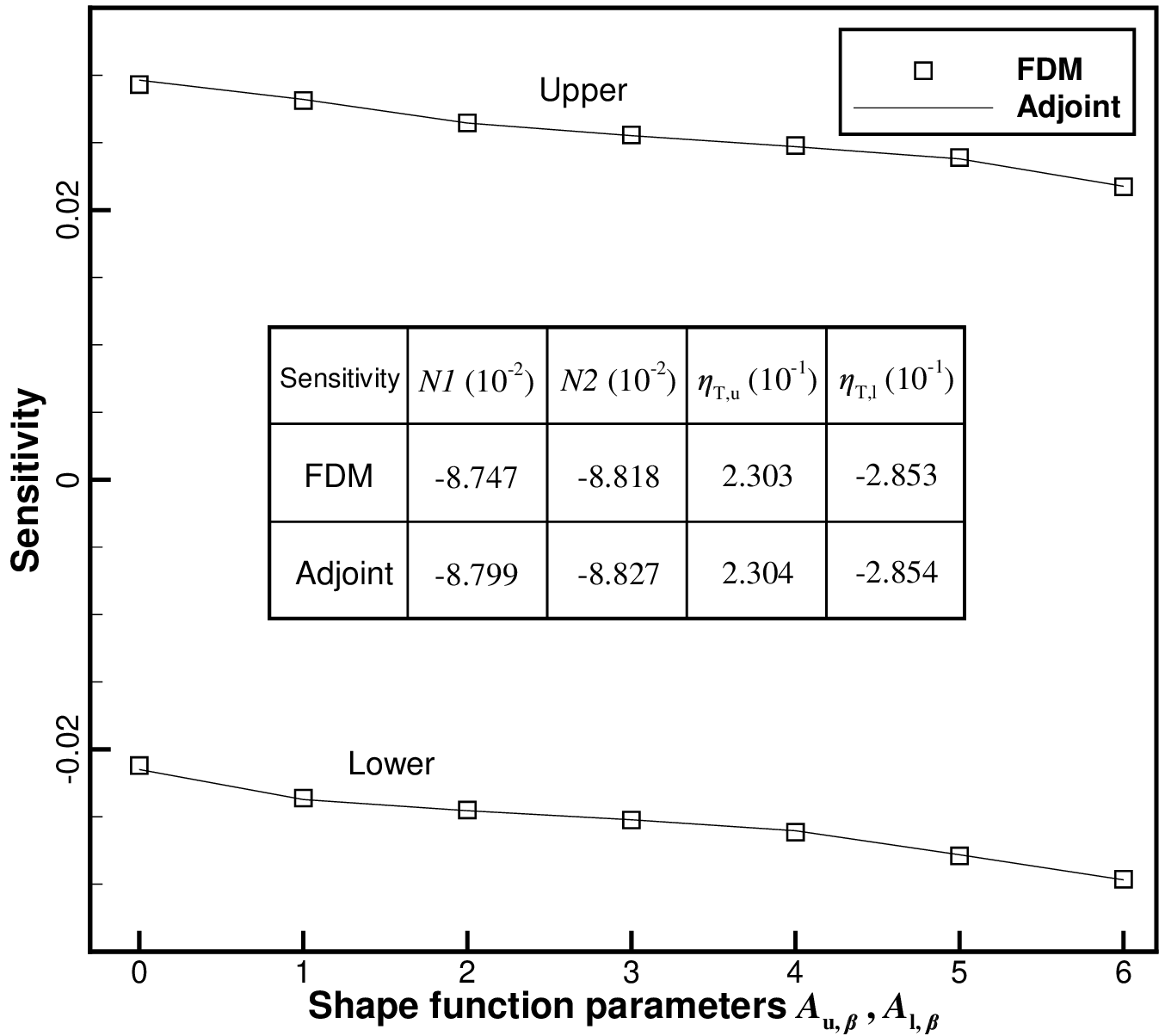}}\hspace{0.02\textwidth}
\subfigure[${\rm Kn}=0.001$]{\includegraphics[width=0.47\textwidth]{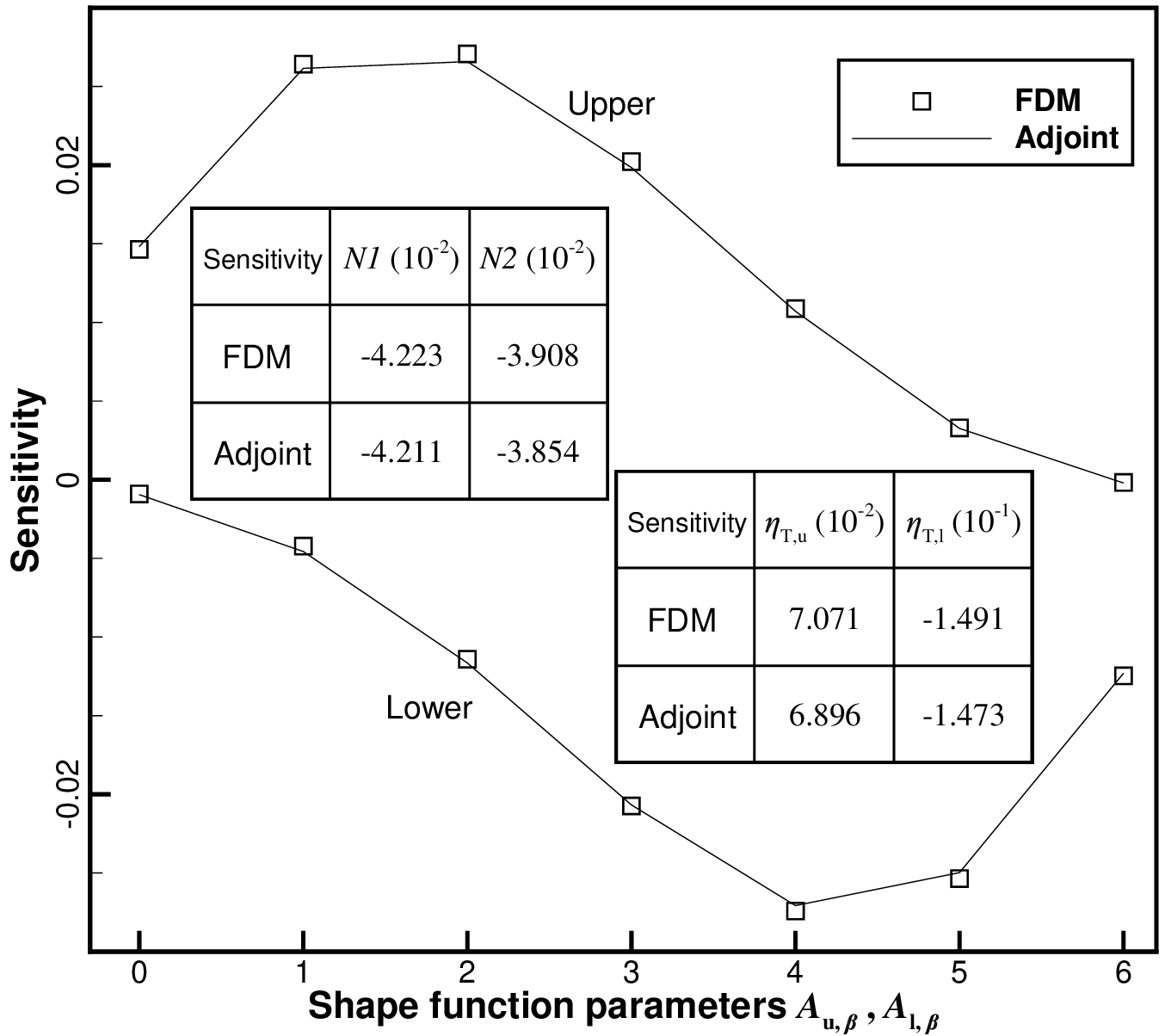}}
\caption{\label{fig:case0_kn01001_senscst} Flow over an elliptic cylinder inside a channel: sensitivity of the drag with respect to the CST design variables. Computed by our method and FDM.}
\end{figure}

When ${\rm Kn}=0.1$ and 0.001, the comparison between our adjoint method and FDM is carried out in Fig.~\ref{fig:case0_kn01001_senscst}. Because the profiles of sensitivity with respect to the point coordinates also oscillate under these two conditions, suffering the problem similar to that in the case of ${\rm Kn}=10$, here we only show the sensitivity with respect to the CST design variables. $30\times30$ and $20\times20$ uniform discretizations of molecular velocity are employed for  ${\rm Kn}=0.1$ and ${\rm Kn}=0.001$ respectively, and they all meet the mesh independence for velocity space. It is shown that the our adjoint results agree well with the FDM results. In conclusion, our method can predict accurate sensitivity for both rarefied and continuum gas flows.

\subsection{Optimization of the airfoil: low-speed cases}\label{sec:case1}

\begin{figure}
\centering
\subfigure[Mesh around the airfoil]{\includegraphics[width=0.47\textwidth]{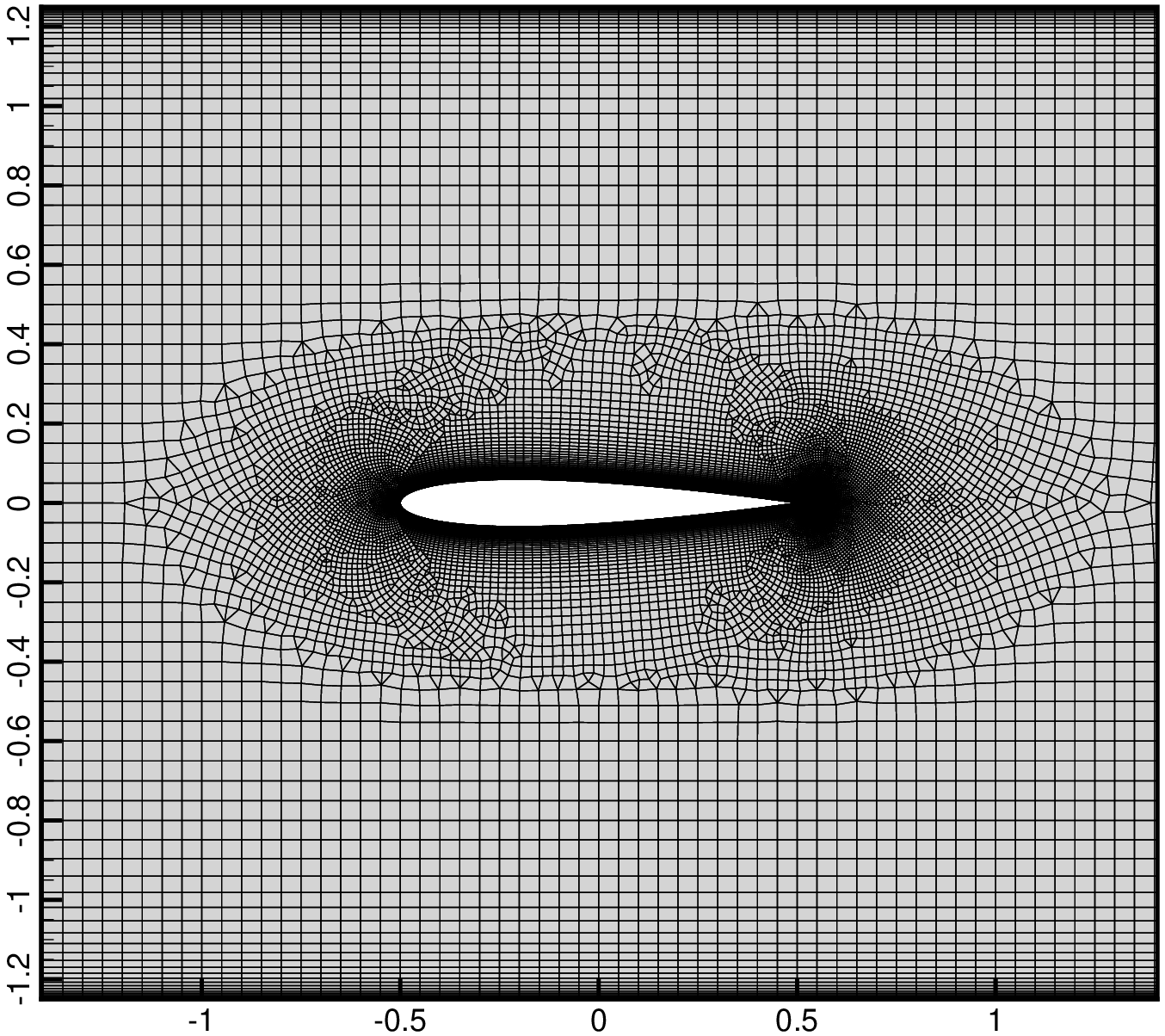}}\hspace{0.02\textwidth}
\subfigure[Mesh near the trailing edge]{\includegraphics[width=0.47\textwidth]{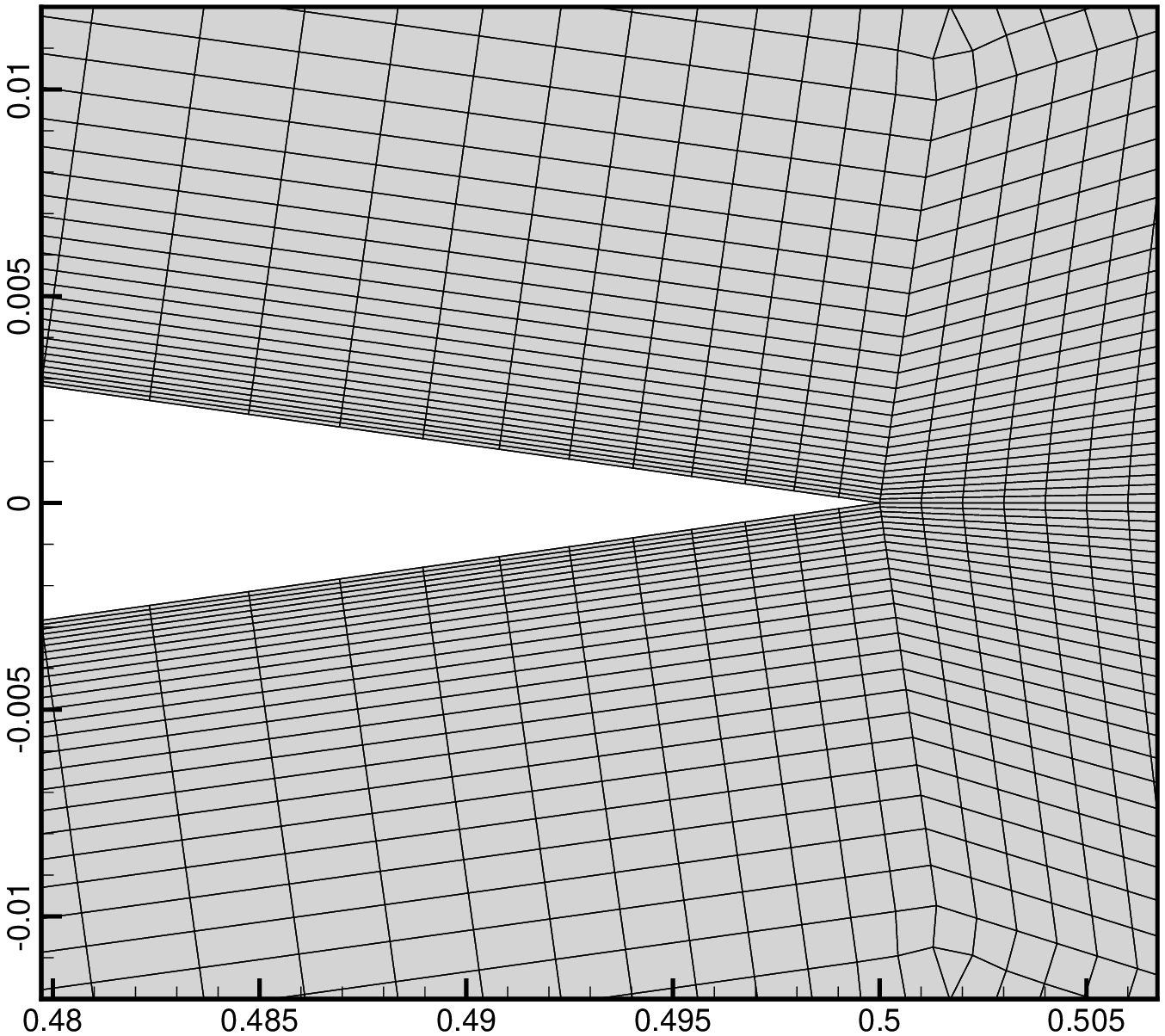}}
\caption{\label{fig:case1_mesh} Optimization of the airfoil inside a channel: the mesh for the initial NACA0012 airfoil, ${\rm Kn}=0.001$.}
\end{figure}

To verify the whole procedure of the present shape optimization method presented in Section \ref{sec:method_optframework}, the shape of the airfoil under the subsonic flow inside a channel is optimized for drag reduction.
The setups of the channel, including the geometry and the flow/boundary conditions, are the same as those in Section \ref{sec:case0}. The initial shape of the airfoil is set as the NACA0012 with a sharp trailing edge, whose exact definition  can be found in Ref.~ \cite{d2naca0012}. The airfoil is centered at $(0,0)^\top$ with $0^\circ$ angle of attack, and has a unit chord length of $c=1$.
The objective is the drag exerted on the airfoil, namely the objective function is $J$ in Eq.~\eqref{eqn:formula_opt} with $m_{J}=v_1$.
The volume constraint in Eq.~\eqref{eqn:formula_opt} is imposed, where the minimum area $V_{\rm min}$ is set as the initial area of the NACA0012 airfoil. In addition, the following constraints are enforced to avoid non-physical geometries and design variables:
\begin{equation}\label{eqn:cst_constraint}
\left.
\begin{aligned}
& N1 \ge 0.01,\quad {A_{{\rm{u}},\beta }} \ge 0,\quad {\eta _{{\rm{T}},{\rm{u}}}} = 0,\\
& N2 \ge 0.01,\quad {A_{{\rm{l}},\beta }} \le 0,\quad {\eta _{{\rm{T}},{\rm{l}}}} = 0.
\end{aligned}
\right\}
\end{equation}
Note that here we force a sharp trailing edge to facilitate the mesh treatment around it.
Three degrees of gas rarefaction are considered from free-molecular to continuum regimes: ${\rm Kn}=10,0.1,0.001$, where Kn is defined with the chord length $c$.
For the discretization of the physical space, two sets of non-uniform unstructured meshes with 28404 cells and 39218 cells in total are adopted for ${\rm Kn}=10,0.1$ and ${\rm Kn}=0.001$ respectively, where the height of the first layer of the mesh adjacent to the airfoil is refined to $10^{-3}c$ and $10^{-4}c$ respectively. Figure~\ref{fig:case1_mesh} shows some details of the initial mesh near the airfoil.
For the discretization of the velocity space, $60 \times 60, 30\times30$ and $20\times20$ uniform meshes in the velocity range $[ - 6{a_\infty },6{a_\infty }]$ are employed, where $a_\infty$ is the far-field acoustic speed. We have ensured that the mesh independence is properly achieved for both the physical space and velocity space under different flow conditions.

\begin{figure}[p]
	\centering
	{\subfigure[${\rm Kn}=10$]{%
			\includegraphics[width=0.47\textwidth]{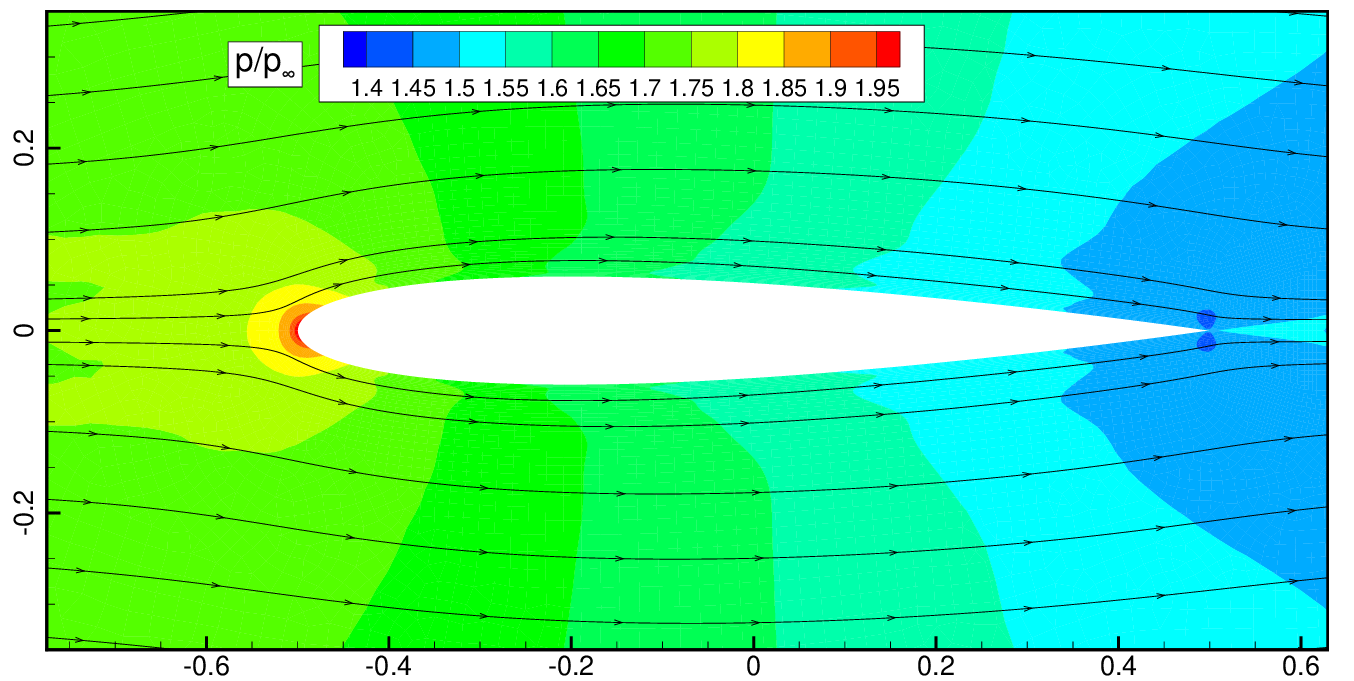}\hspace{0.02\textwidth}%
			\includegraphics[width=0.47\textwidth]{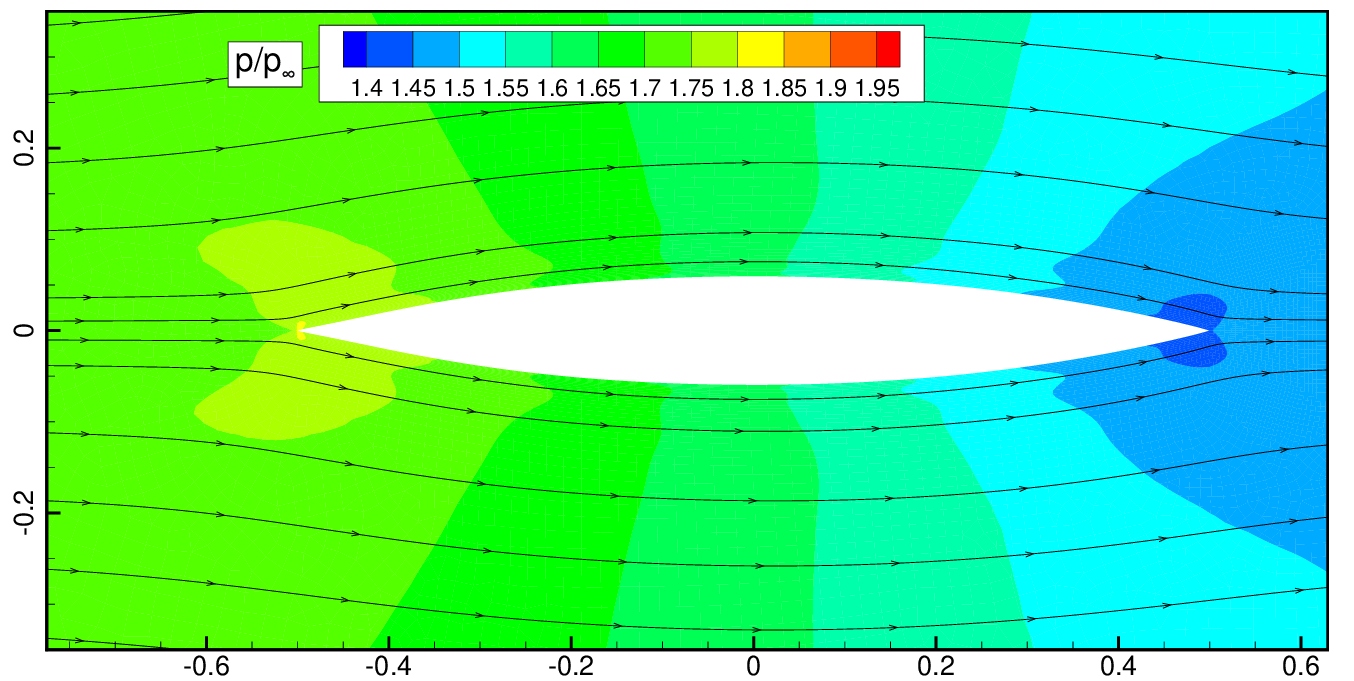}}}\\%
	{\subfigure[${\rm Kn}=0.1$]{%
			\includegraphics[width=0.47\textwidth]{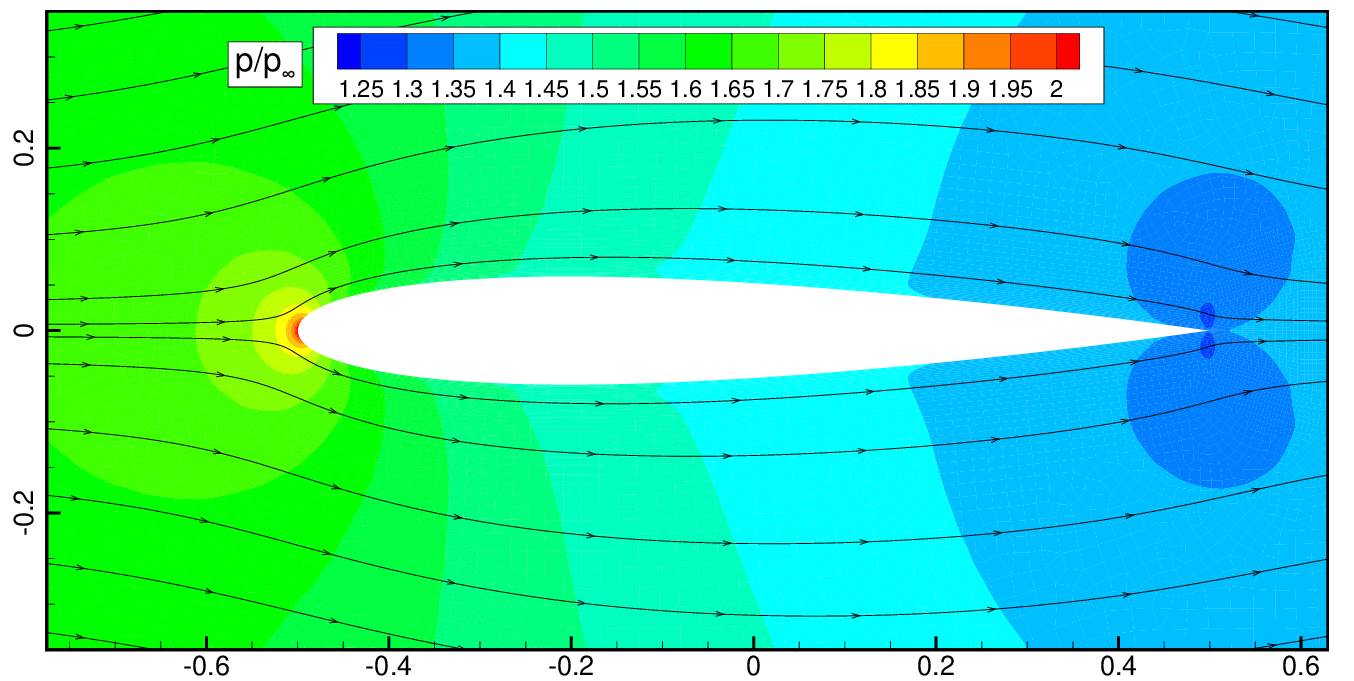}\hspace{0.02\textwidth}%
			\includegraphics[width=0.47\textwidth]{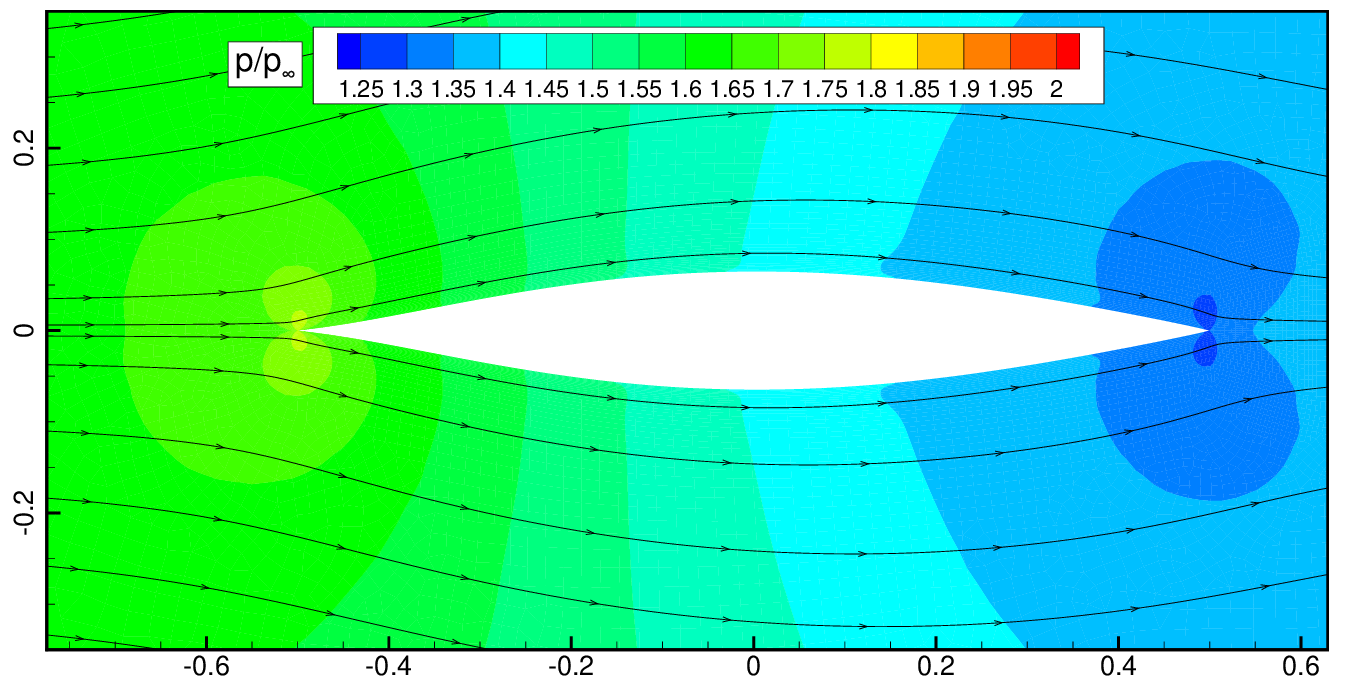}}}\\%
	{\subfigure[${\rm Kn}=0.001$]{%
			\includegraphics[width=0.47\textwidth]{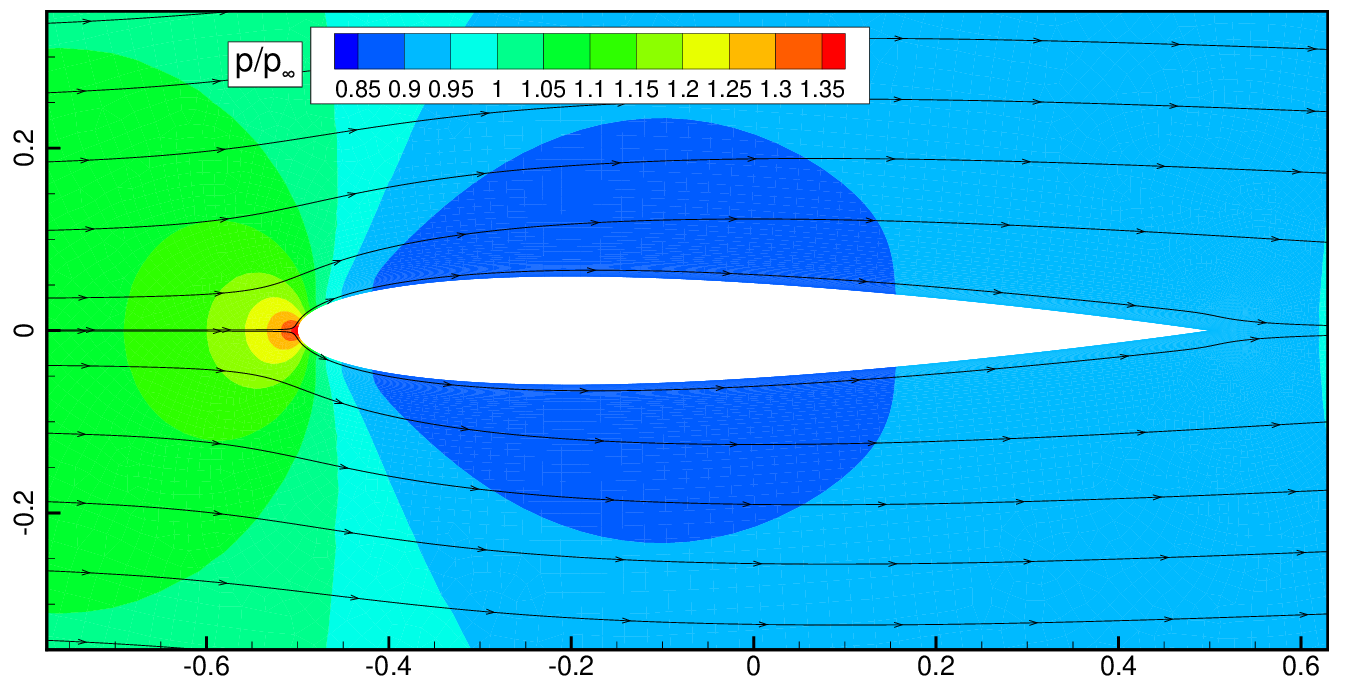}\hspace{0.02\textwidth}%
			\includegraphics[width=0.47\textwidth]{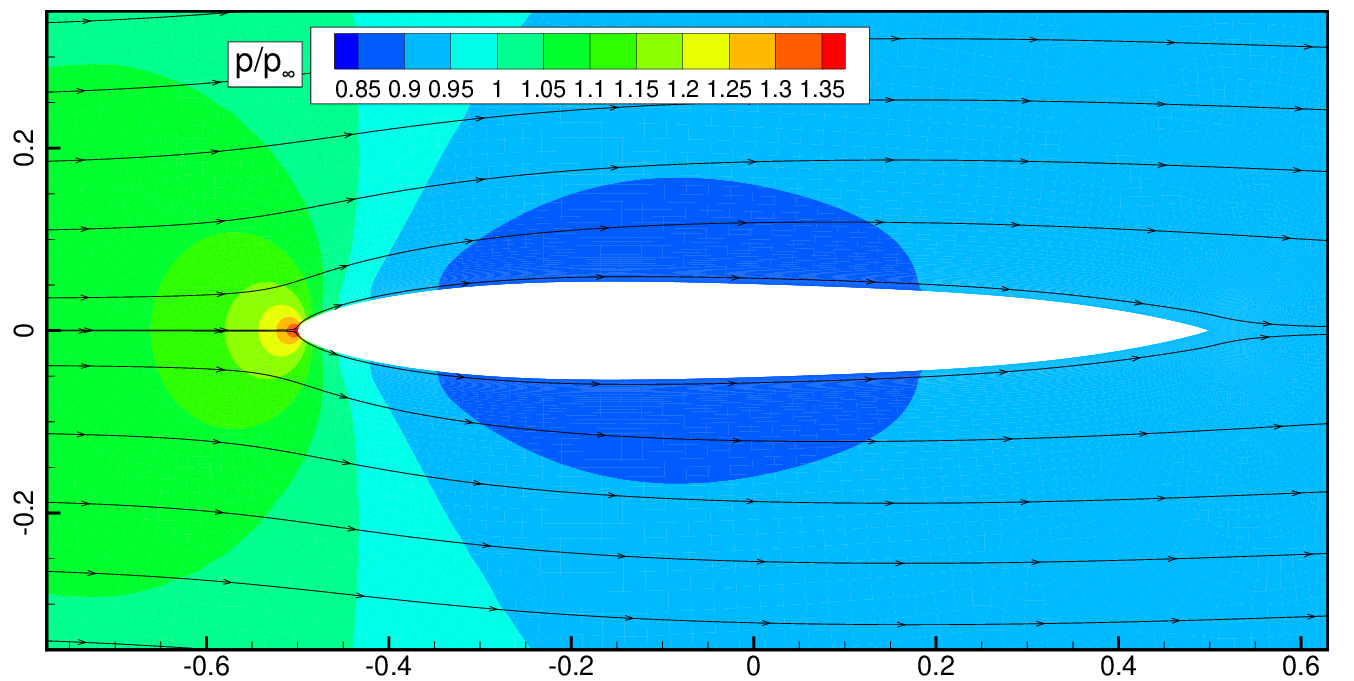}}}
	\caption{\label{fig:case1_pre} Optimization of the airfoil inside a channel: streamlines and pressure distributions before (left) and after (right) optimization.}
\end{figure}

The optimizations follow the procedure described in Section \ref{sec:method_optframework}. The streamlines and pressure distributions around the initial airfoil and the optimized airfoils are shown in Fig.~\ref{fig:case1_pre}. In the figures we can find that:
\begin{enumerate}
\item Under all three conditions, the high-pressure zone in front of the leading edge is weakened after optimization.
\item Under the rarefied conditions of ${\rm Kn}=10,0.1$, the optimized airfoil has a sharp leading edge, decreasing the high pressure on the leading edge effectively.
\item Under the continuum condition ${\rm Kn}=0.001$, the optimized airfoil has the minimal change relative to the initial airfoil among these three conditions. Nevertheless, the optimized airfoil becomes thinner and has more uniform thickness distribution to decrease the pressure drag.
\end{enumerate}

\begin{figure}
\centering
\includegraphics[width=0.47\textwidth]{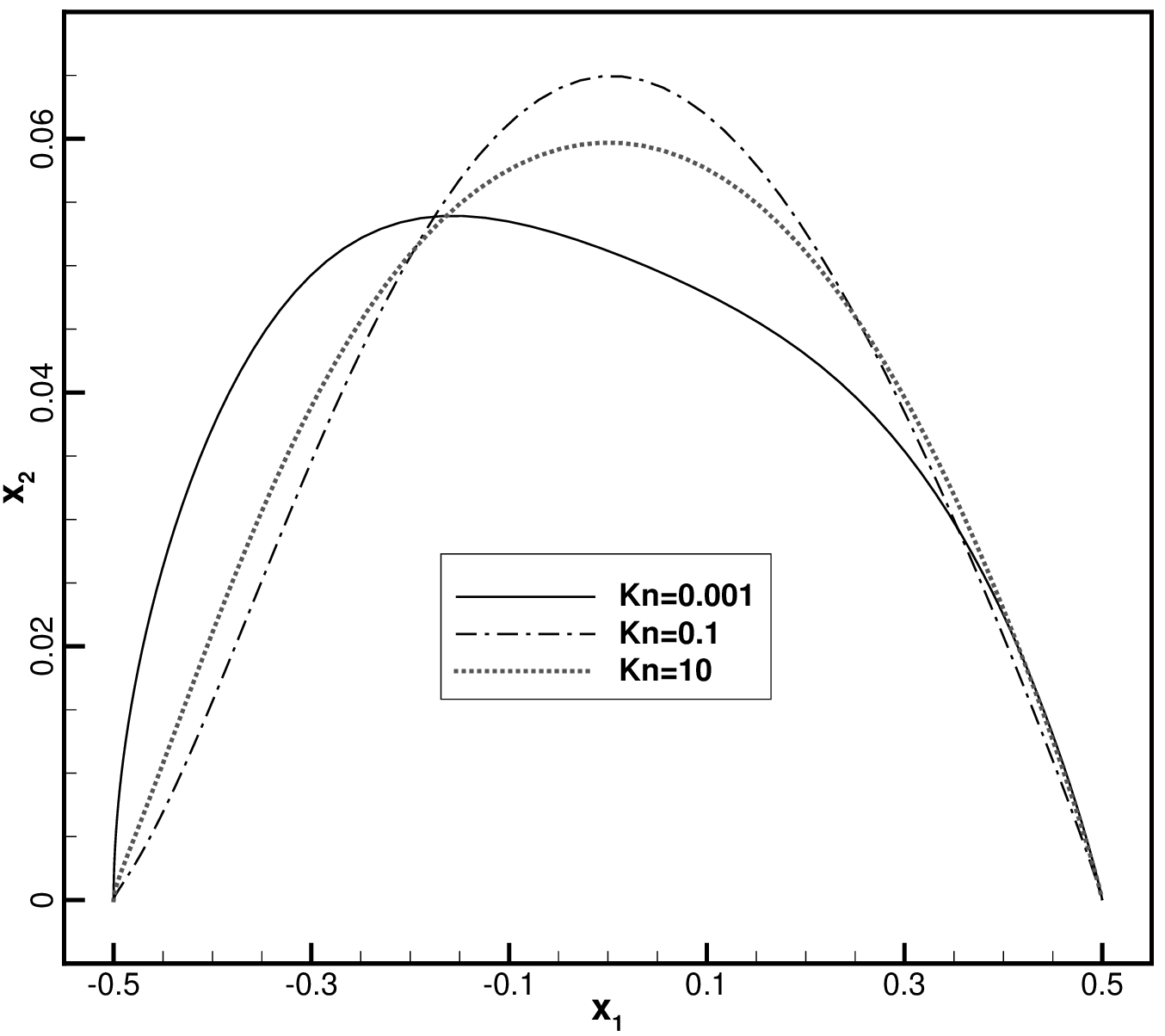}
\caption{\label{fig:case1_suf} Optimization of the airfoil inside a channel: comparison of the shapes of optimized airfoils under different degrees of gas rarefaction.}
\end{figure}

A detailed comparison of the shapes of the optimized airfoils is shown in Fig.~\ref{fig:case1_suf}. The most interesting finding is that the variation of the optimized airfoil thickness with the Knudsen number is not monotonic, and the optimized airfoil for ${\rm Kn}=0.1$ has the maximum thickness among the three optimized airfoils. This trend is in consistent with  the airfoil optimization conducted in our previous work via topology optimization \cite{yuan2024design}. It  can also be found that, for the airfoil optimized in continuum gas flow, the maximum thickness position is closer to the leading edge, while for the airfoil optimized for rarefied gas flow, the maximum thickness location is near the middle of the airfoil.

Table \ref{tab:case1_cdeff} shows the drag coefficients $C_{\rm d}$ of the initial and optimized airfoils. The drag decreases are not so high because the initial NACA0012 airfoil itself has a low drag due to its long thin configuration. Especially, for the case ${\rm Kn}=0.001$ the optimized airfoil has roughly similar shape to the initial one, and the drag reduction is only $1.19\%$. Another interesting thing is that the variation of $C_{\rm d}$ with Kn is also not monotonic, and the maximum $C_{\rm d}$ occurs at ${\rm Kn}=0.1$ among these three flow conditions, which can be hardly explained by the linear constitutive relation.

The efficiency of the method is also shown in Table \ref{tab:case1_cdeff}. All computations adopt the parallel computing and are conducted on the cluster with ``\emph{Intel(R) Xeon(R) Gold 6148 CPU @ 2.40GHz}'' (40 cores per node). Benefiting from
the high convergence efficiency of the quasi-Newton optimizer and the accurate sensitivity provided by our method, the optimizations generally converge in a dozen steps and finish in 6-20 minutes. The case of ${\rm Kn}=10$ consumes more computational resources because of the large number of velocity points required. Generally speaking, as a shape optimization method capable of addressing the full spectrum of gas rarefaction, the present approach achieves highly desirable efficiency.

\begin{table}[]
	\centering
	\caption{\label{tab:case1_cdeff}Optimization of the airfoil inside a channel: drag coefficients before/after optimization and the optimization efficiency. $C_{\rm d}$ is normalized by the far-field state $\frac{1}{2}\rho _\infty\bm u_\infty^2c$.}\vspace{0.3cm}
\begin{tabular}{ccccccccc}
\hline
\multirow{2}{*}{Kn} & \multicolumn{2}{c}{$C_{\rm d}$}                                      & \multirow{2}{*}{\begin{tabular}[c]{@{}c@{}}Drag\\ decrease\end{tabular}} & \multirow{2}{*}{\begin{tabular}[c]{@{}c@{}}Physical\\ mesh\end{tabular}} & \multirow{2}{*}{\begin{tabular}[c]{@{}c@{}}Velocity\\ mesh\end{tabular}} & \multirow{2}{*}{\begin{tabular}[c]{@{}c@{}}Parallel\\ cores\end{tabular}} & \multirow{2}{*}{\begin{tabular}[c]{@{}c@{}}Optim.\\ steps\end{tabular}} & \multirow{2}{*}{\begin{tabular}[c]{@{}c@{}}Time\\ (s)\end{tabular}} \\ \cline{2-3}
                      & \multicolumn{1}{l}{Initial} & \multicolumn{1}{l}{Optimized} &                                                                          &                                                                          &                                                                          &                                                                           &                                                                               &                                                                     \\ \hline
0.001              & 0.1521                      & 0.1502                        & 1.19\%                                                                     & 39218                                                                    & 20×20                                                                    & 40                                                                        & 18                                                                            & 713                                                                 \\
0.1                & 0.9961                      & 0.9710                        & 2.52\%                                                                                                                                         & 28404                                                                    & 30×30                                                                    & 80                                                                        & 9                                                                             & 390                                                                 \\
10                 & 0.6885                      & 0.6476                        & 5.95\%                                                                                                                                         & 28404                                                                    & 60×60                                                                    & 160                                                                       & 12                                                                            & 1133                                                                \\ \hline
\end{tabular}
\end{table}



\begin{figure}
	\centering
	{\subfigure[${\rm Kn}=10$]{%
			\includegraphics[width=0.47\textwidth]{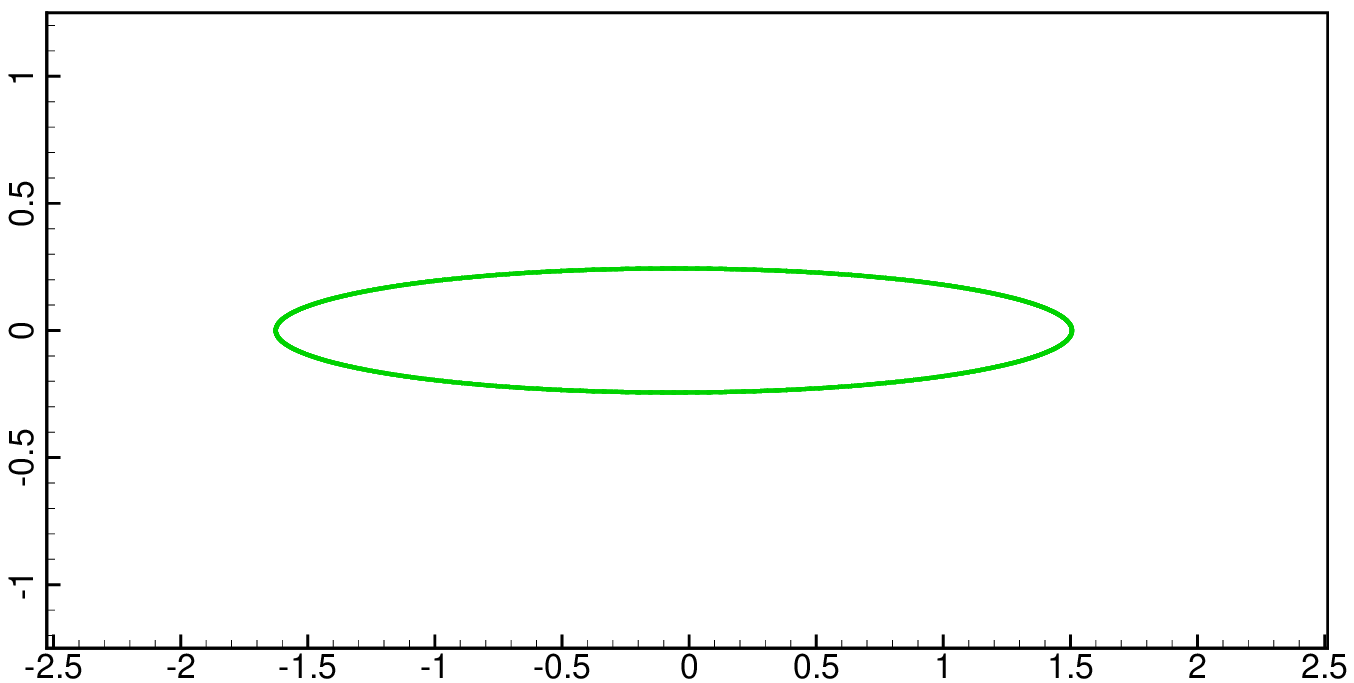}\hspace{0.02\textwidth}%
			\includegraphics[width=0.47\textwidth]{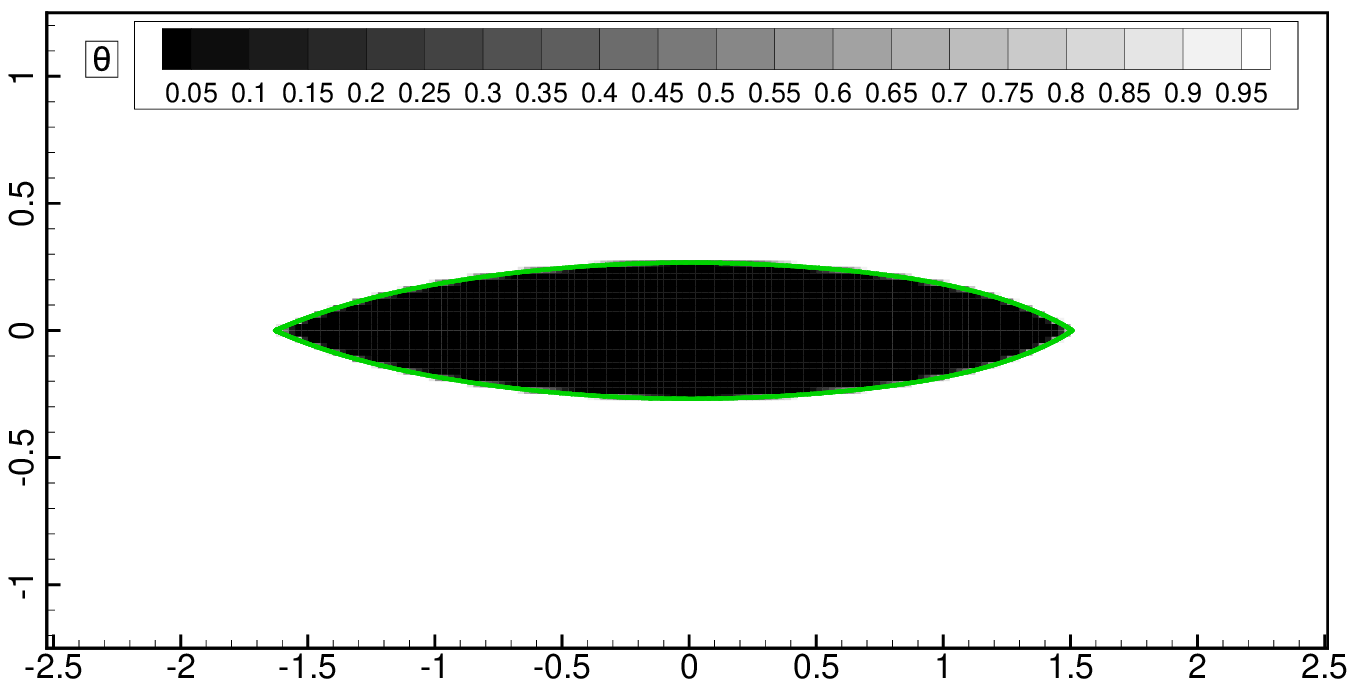}}}\\%
	{\subfigure[${\rm Kn}=0.5$]{%
			\includegraphics[width=0.47\textwidth]{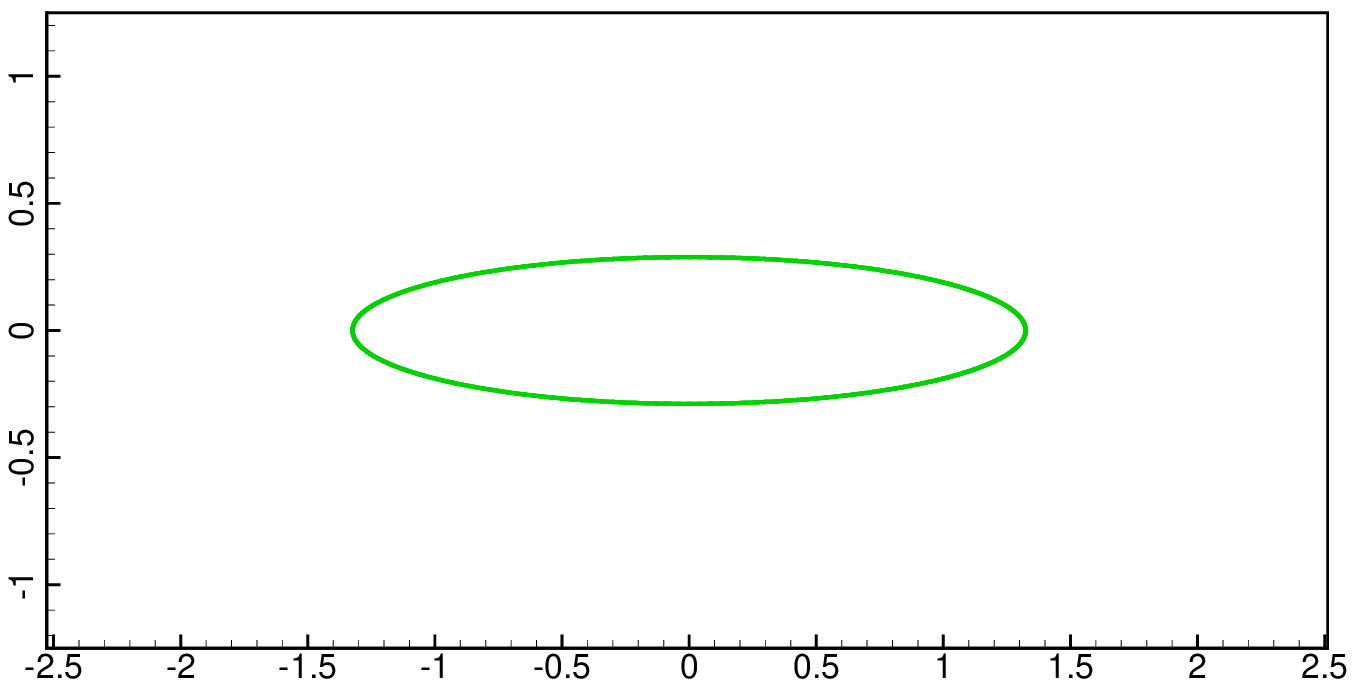}\hspace{0.02\textwidth}%
			\includegraphics[width=0.47\textwidth]{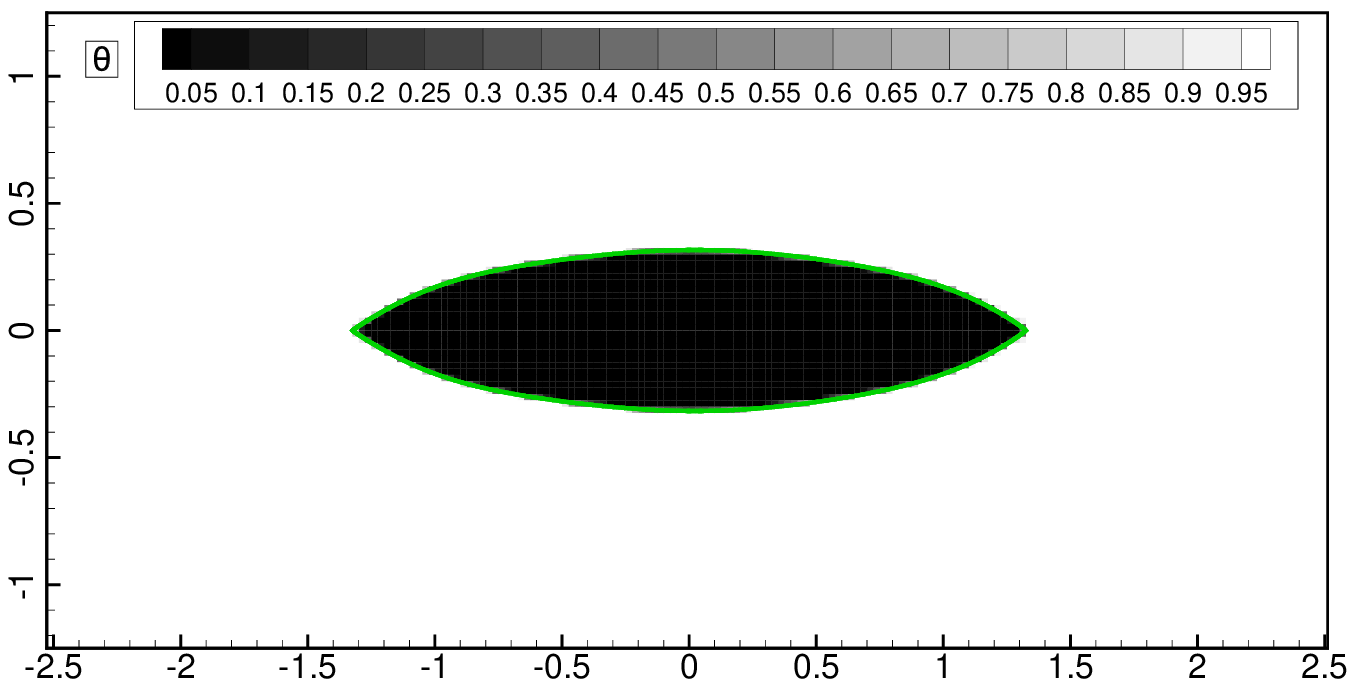}}}\\%
	{\subfigure[${\rm Re}=200$]{%
			\includegraphics[width=0.47\textwidth]{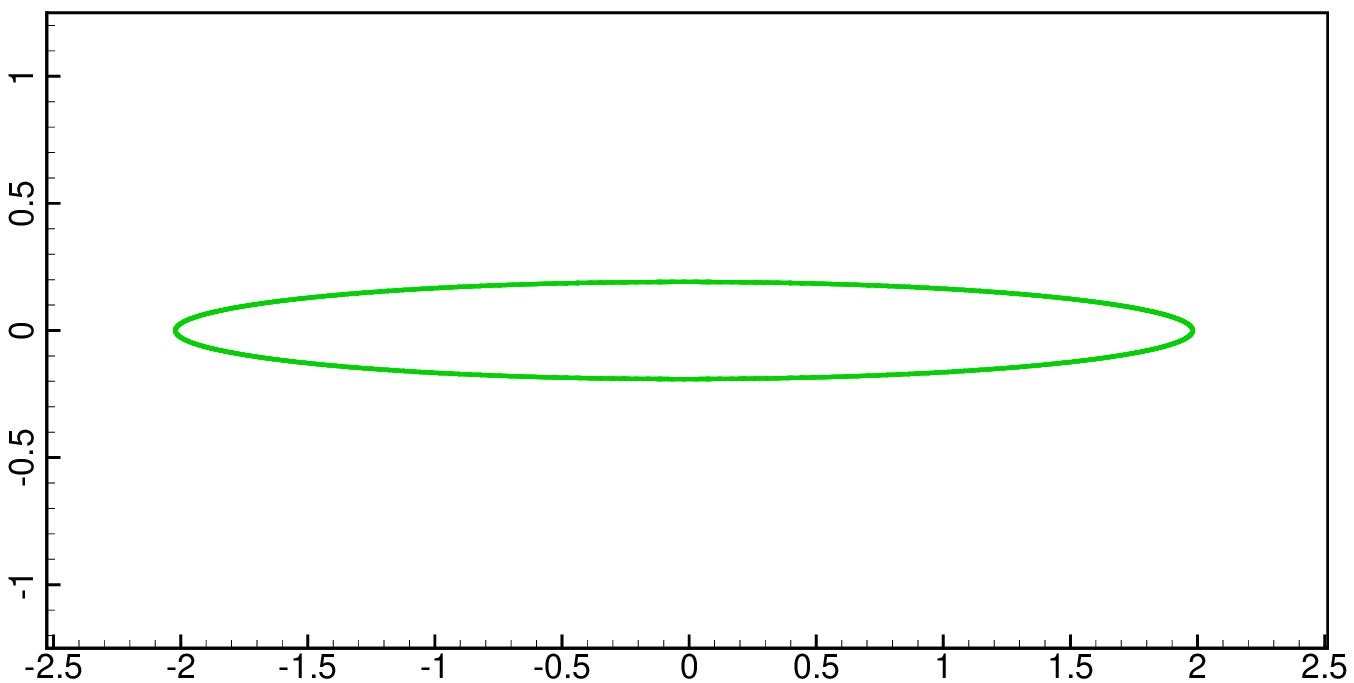}\hspace{0.02\textwidth}%
			\includegraphics[width=0.47\textwidth]{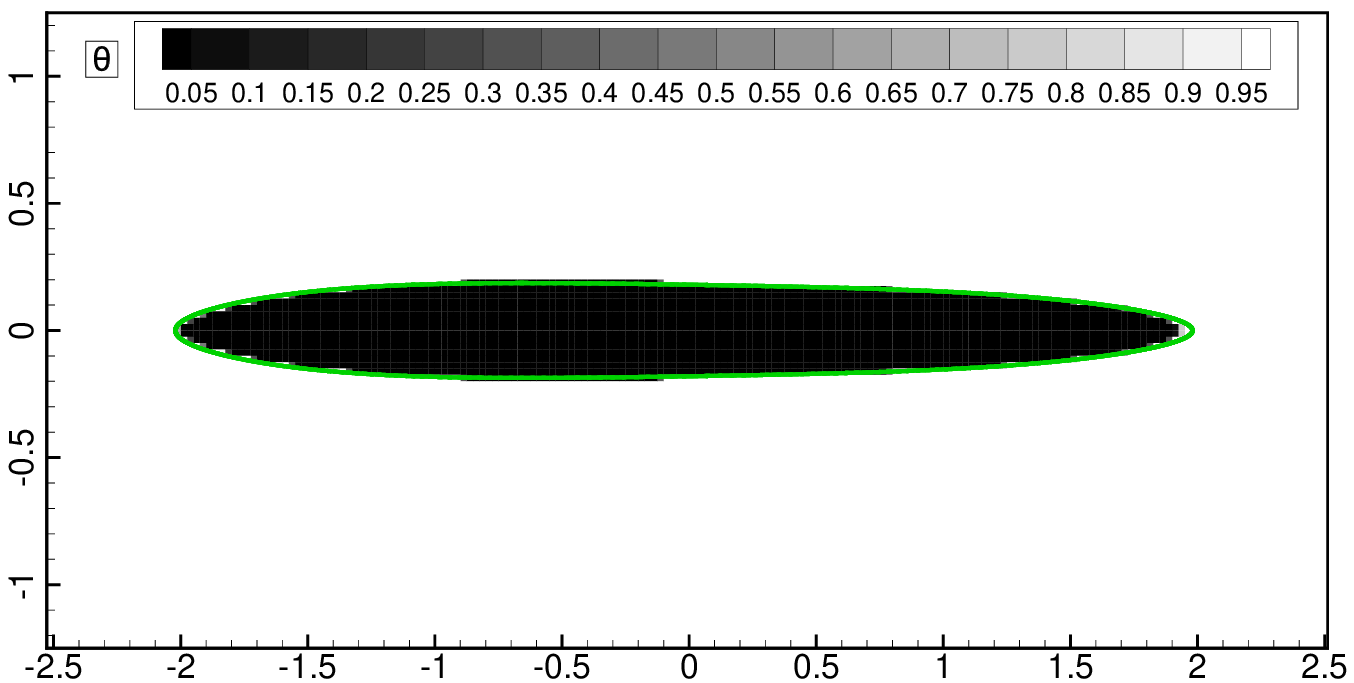}}}
	\caption{\label{fig:case1_topo} Optimization of the airfoil inside a channel: the comparison of the results calculated by the present shape optimization method (green lines) with those obtained by the topology optimization method \cite{yuan2024design} (grayscale fields). The left column is the initial setup for the present method. Kn and Re are defined by a reference length of 2 (completely same condition with Ref. \cite{yuan2024design}). The result of topology optimization for ${\rm Re}=200$ is different from the published data of Ref. \cite{yuan2024design}, this is because the optimization in \cite{yuan2024design} did not fully converge for this condition.}
\end{figure}

To validate the method we have further compared the results of the present shape optimization method with the results obtained by the topology optimization method  \cite{yuan2024design}, through the design problem of the airfoil optimization for drag reduction inside a channel.
Because the CST parameterization adopted here assumes a fixed chord length, while this is not the case in topology optimization, it is not able to reproduce exactly the same optimization problem carried out in Ref. \cite{yuan2024design}. Nevertheless, in the present method we avoid this issue by starting the optimization from the initial shape of an ellipse with the same area-length ratio $V_{\min}/c^2$ with that of the optimized airfoil in \cite{yuan2024design}, namely we directly set the initial chord length to the optimal value.
It is evident from figure \ref{fig:case1_topo} that both types of optimization result in identical optimal airfoil shapes.

\subsection{Optimization of the airfoil: high-speed cases}\label{sec:case2}

\begin{figure}[t]
	\centering
	\subfigure[Global view]{\includegraphics[width=0.47\textwidth]{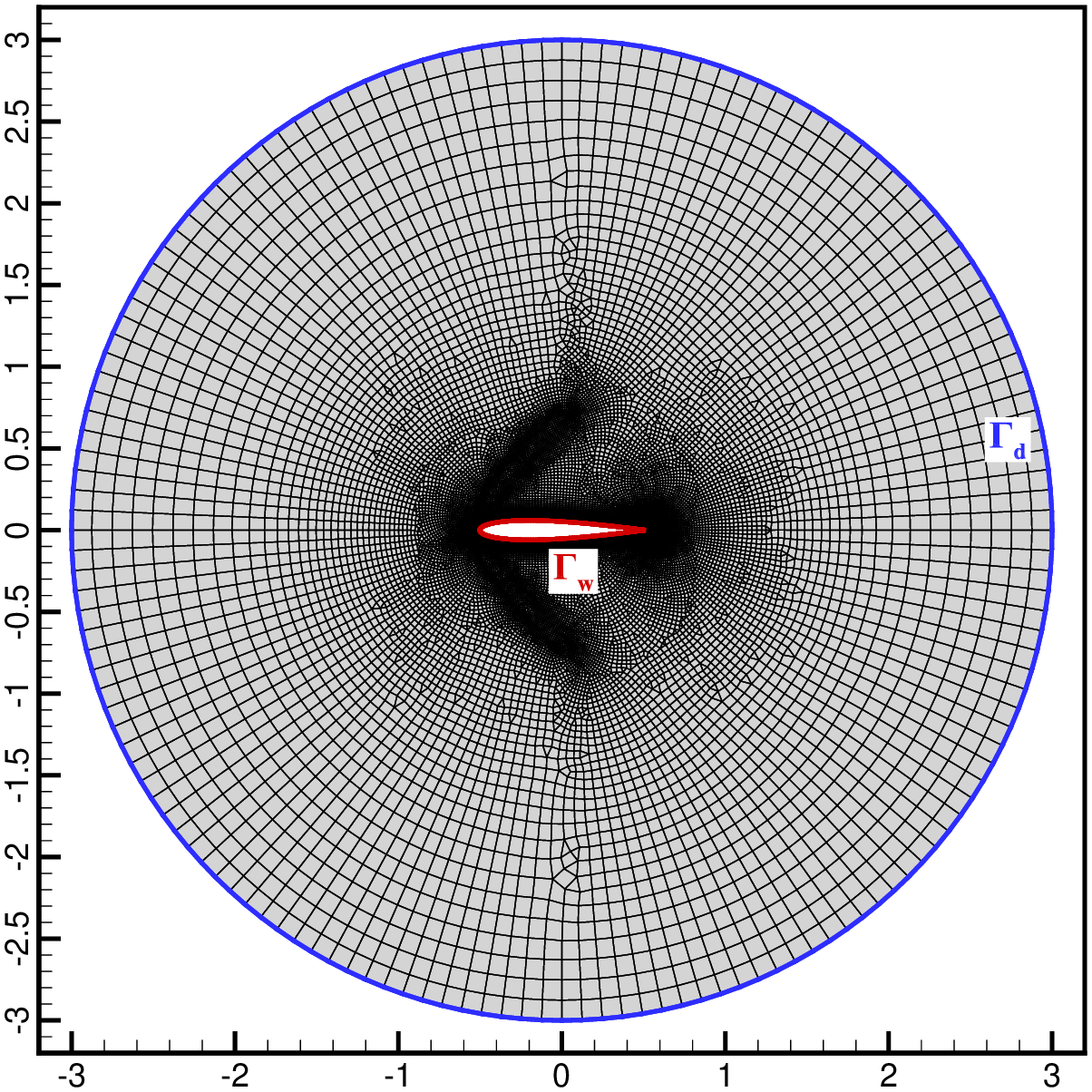}}\hspace{0.02\textwidth}
	\subfigure[Mesh near the airfoil]{\includegraphics[width=0.47\textwidth]{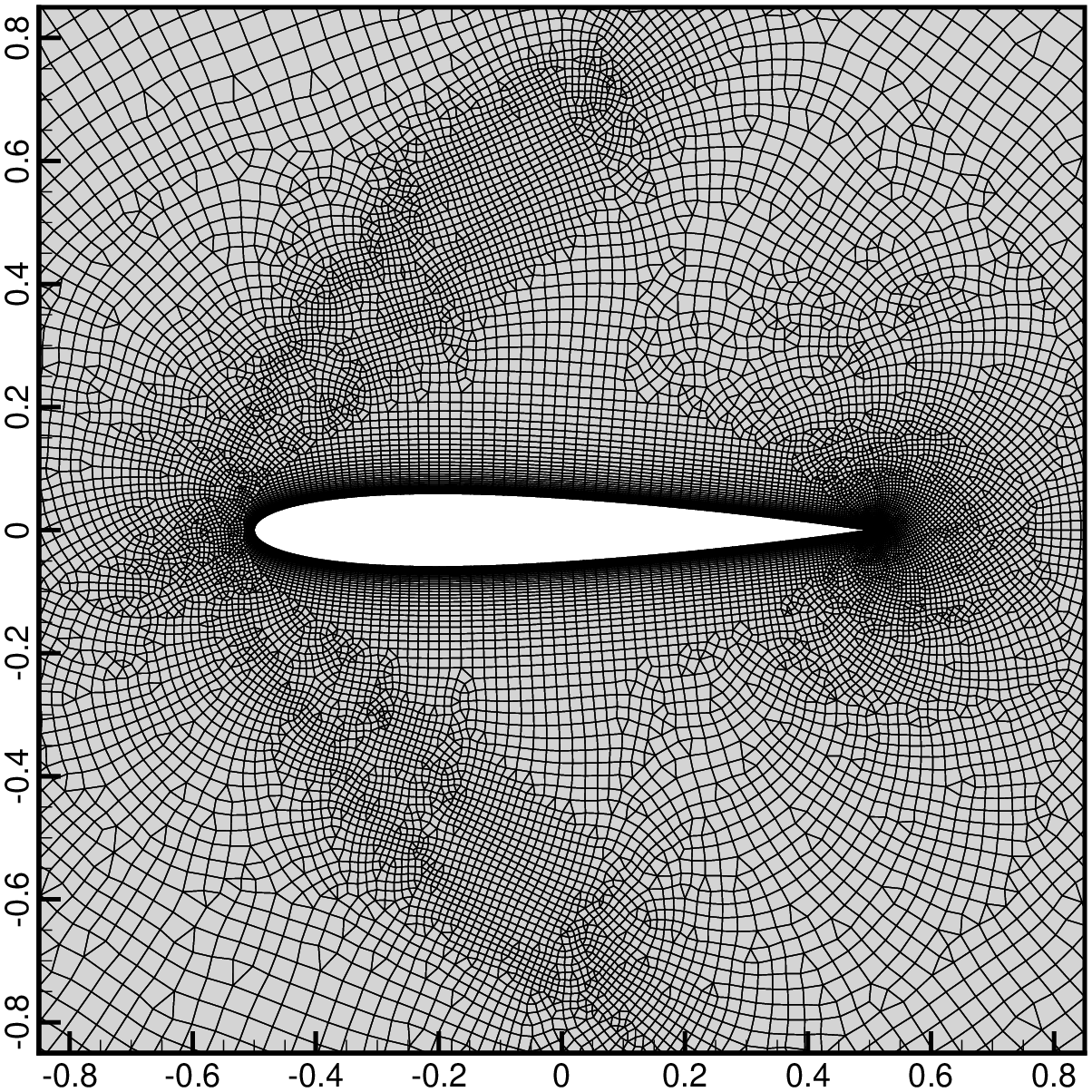}}
	\caption{\label{fig:case2_mesh} Optimization of the airfoil under supersonic flow: the boundary setup and the mesh for the initial NACA0012 airfoil.}
\end{figure}

To test the performance of our method for high-speed cases, we carry out the drag-reduction optimization of the airfoil under supersonic flow in this section. The Mach number is ${\rm Ma}=2$, and two degrees of rarefaction, ${\rm Kn}=0.01, 0.5$ relative to the chord length, are considered. 
The setup for the computational domain is shown in Fig.~\ref{fig:case2_mesh}. At the far-field boundary $\Gamma _{\rm d}$ the Dirichlet boundary condition of Eq.~\eqref{eqn:drlt_maxwell} is imposed, where the gas state corresponds to the free-stream condition of ${\rm Ma}=2$. On the solid wall $\Gamma _{\rm w}$, the diffuse boundary condition of Eq.~\eqref{eqn:formula_bgk_fdw0} is imposed with ${\bm u_{\rm w}} = 0,{T_{\rm w}} = {T_\infty }$. 
The airfoil is centered at $(0,0)^\top$ with zero angle of attack, the chord length is set as $c=1$. For the initial airfoil, similar to Section \ref{sec:case1}, it has a shape of the NACA0012 with sharp trailing edge \cite{d2naca0012}.
The setups for the objective and the volume constraint are the same as those in Section \ref{sec:case1}, and the constraint of Eq.~\eqref{eqn:cst_constraint} to guarantee a physical geometry is also enforced.

\begin{figure}[p]
\centering
\includegraphics[width=0.4\textwidth]{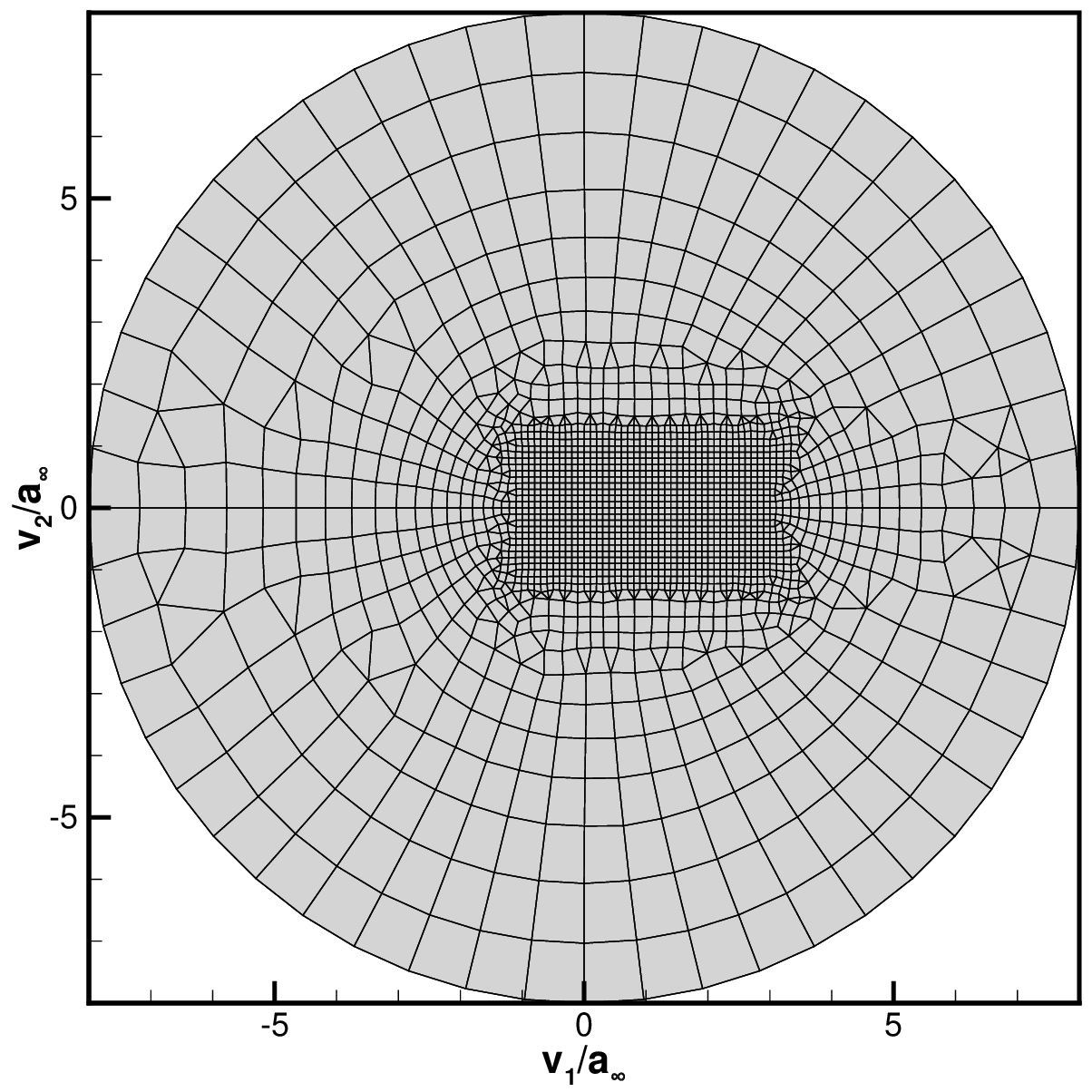}
\caption{\label{fig:case2_meshmic} Optimization of the airfoil under supersonic flow: the mesh of the velocity space employed in the case ${\rm Kn}=0.5$.}
\end{figure}

\begin{figure}[p]
	\centering
	{\subfigure[${\rm Kn}=0.5$]{%
			\includegraphics[width=0.4\textwidth]{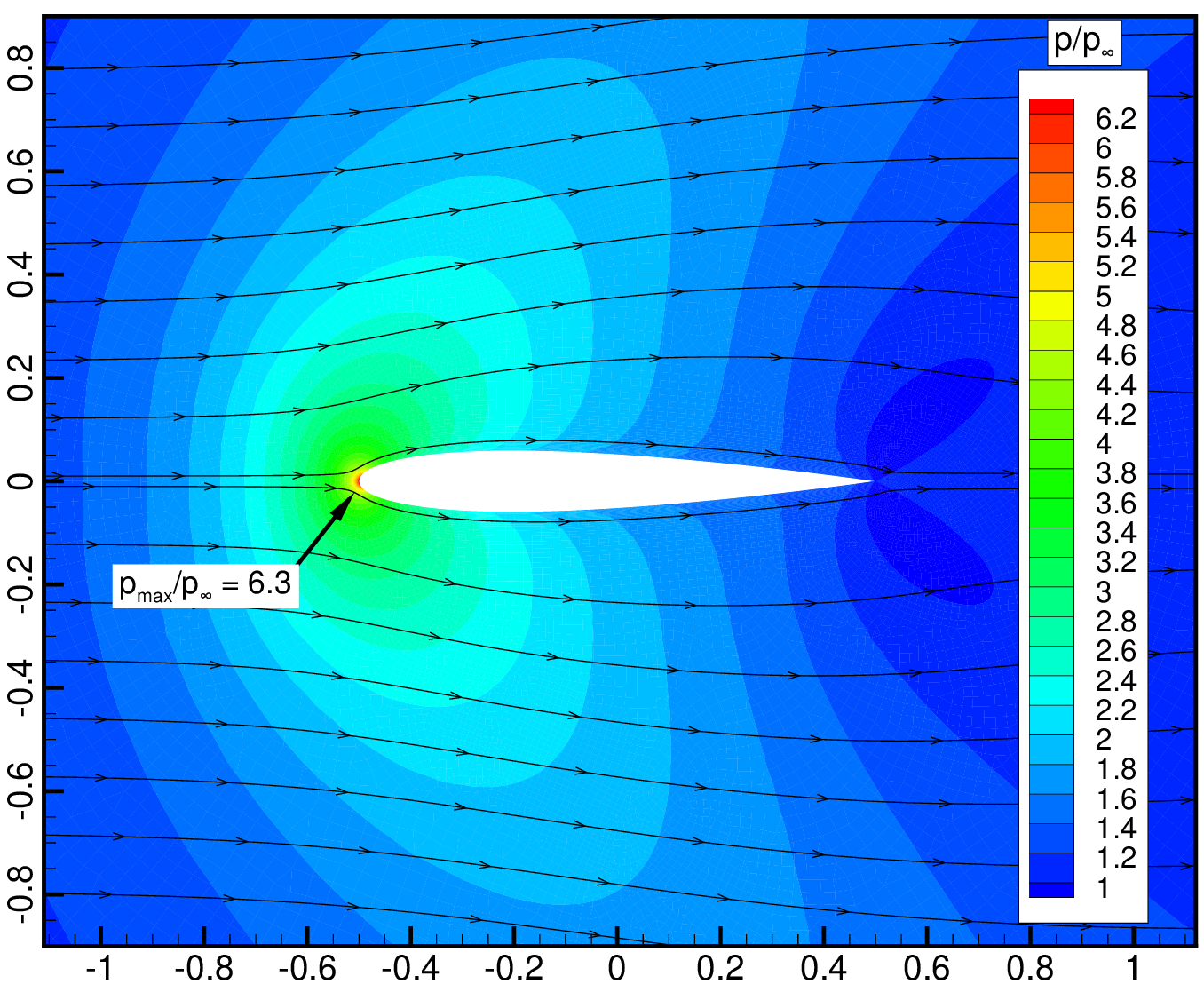}\hspace{0.02\textwidth}%
			\includegraphics[width=0.4\textwidth]{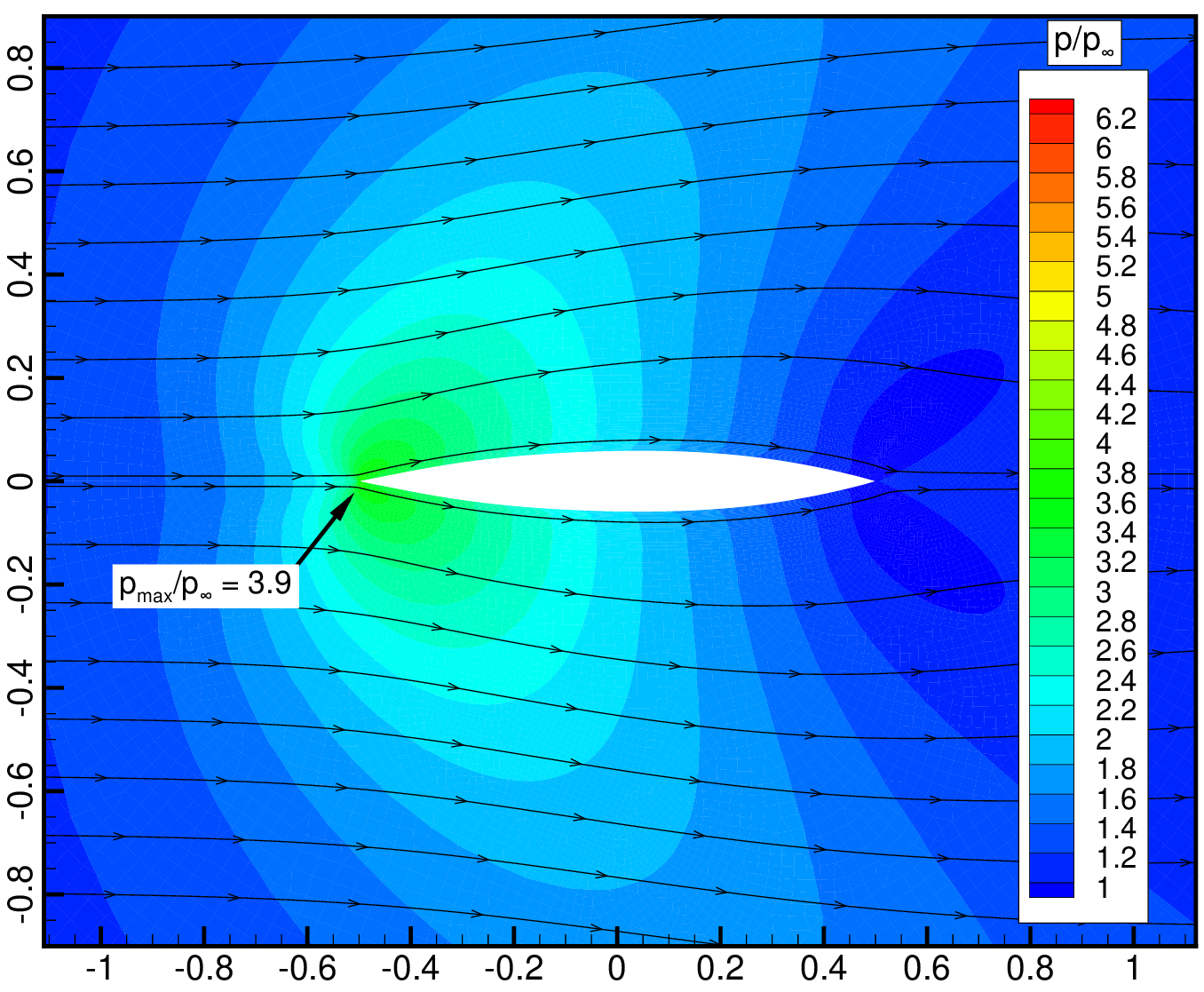}}}\\%
	{\subfigure[${\rm Kn}=0.01$]{%
			\includegraphics[width=0.4\textwidth]{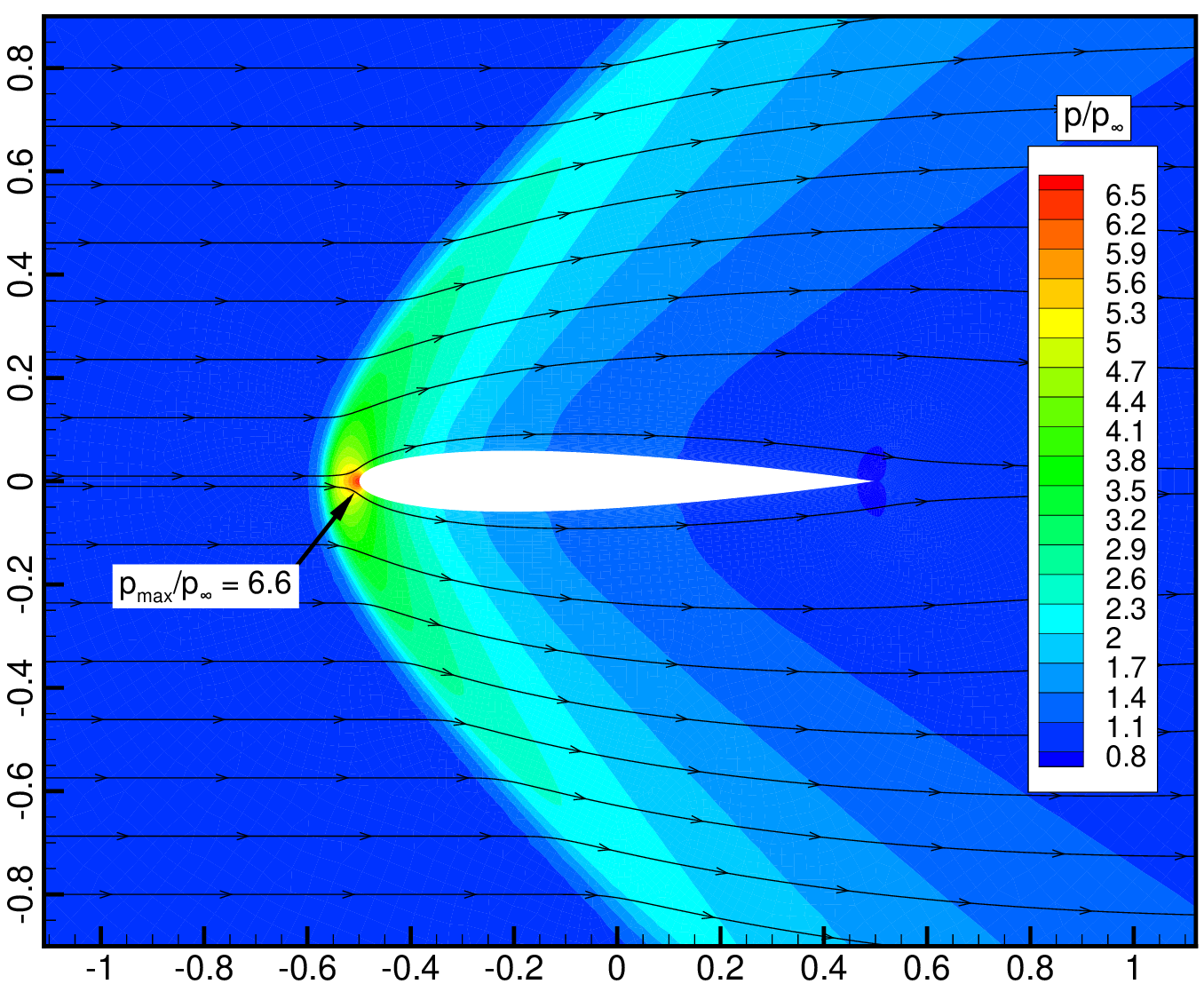}\hspace{0.02\textwidth}%
			\includegraphics[width=0.4\textwidth]{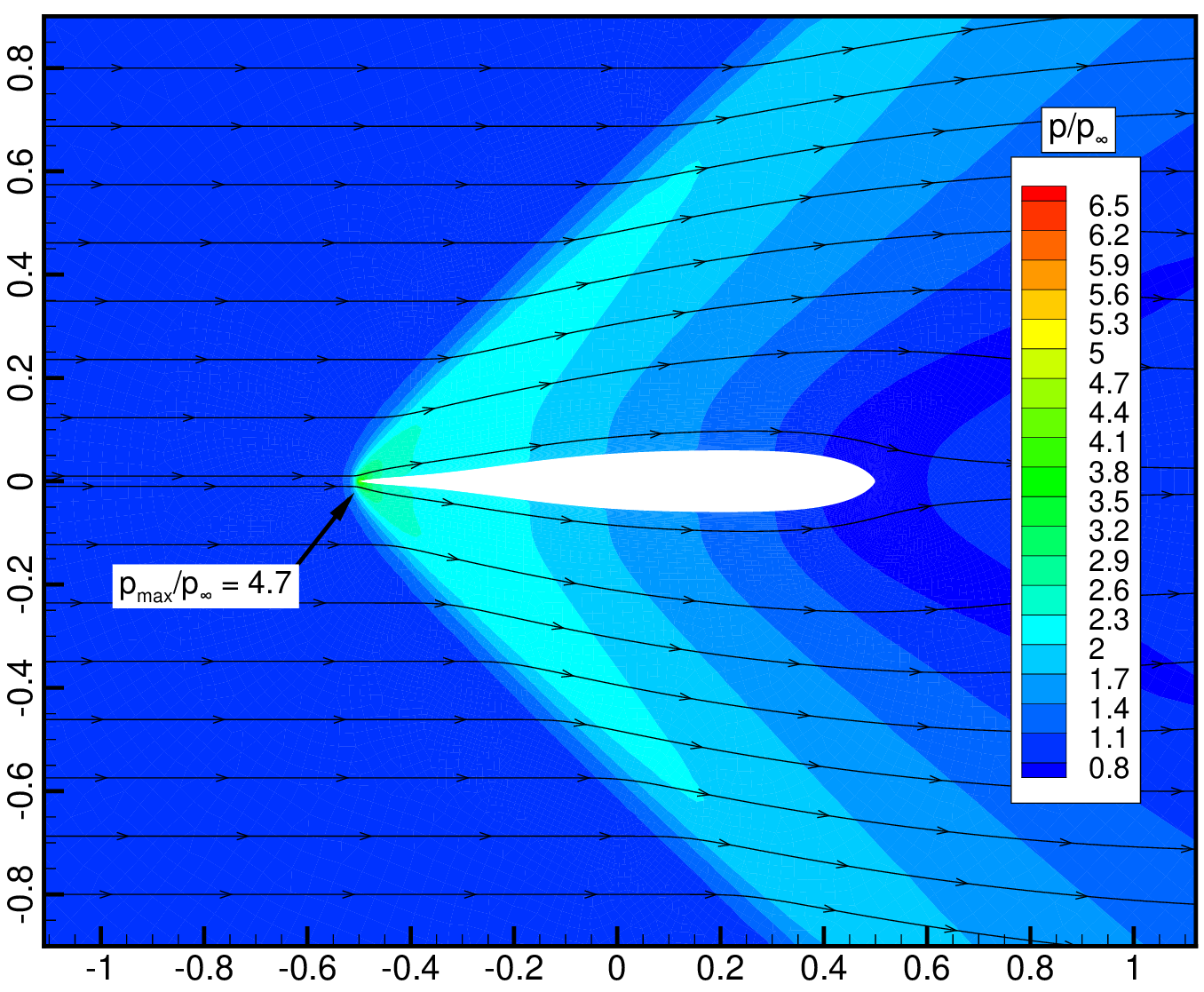}}}
	\caption{\label{fig:case2_pre} Optimization of the airfoil under supersonic flow: streamlines and pressure distributions before (left) and after (right) optimization.}
\end{figure}

For the discretization of the physical space, a non-uniform unstructured mesh with totally 21666 cells is adopted in Fig.~\ref{fig:case2_mesh}. The first layer of the mesh adjacent to the solid wall is refined to have a height of $10^{-3}c$, and the mesh around the region of the shock wave is also refined as shown in the figure.
For the discretization of the velocity space, in the case ${\rm Kn}=0.01$ we employ the $28\times28$ uniform mesh in the velocity range $[ - 8{a_\infty },8{a_\infty }]$ where $a_\infty$ is the free-stream acoustic speed. In the case ${\rm Kn}=0.5$ a non-uniform unstructured mesh with 1984 cells is employed  in Fig.~\ref{fig:case2_meshmic}, where the mesh around the velocity $(0,0)^\top$ (wall speed) and $(2{a_\infty },0)^\top$ (free-stream speed) is refined. More information about the principle of generating the velocity mesh can be found in Refs. \cite{yuan2020conservative,yuan2021application,zhang2025implicit}. 
For all cases the mesh independence has been verified in both physical and velocity spaces. 

\begin{figure}[t]
\centering
\includegraphics[width=0.47\textwidth]{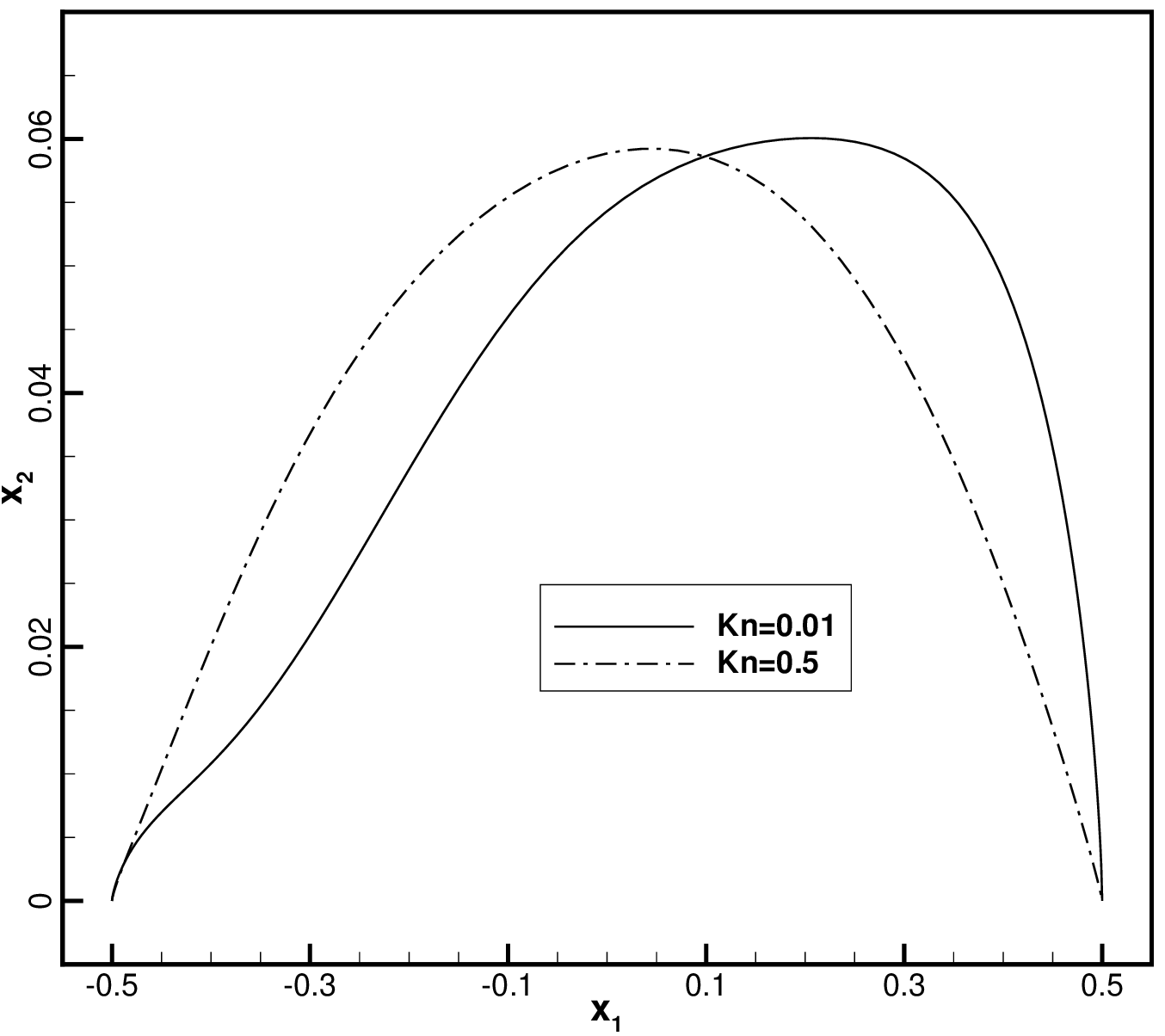}
\caption{\label{fig:case2_suf} Optimization of the airfoil under supersonic flow: comparison of the shapes of the optimized airfoils for different Knudsen numbers.}
\end{figure}

The optimization follows the procedure described in Section \ref{sec:method_optframework}.
The flow fields around the initial and the optimized airfoils are shown in Fig.~\ref{fig:case2_pre}. 
In both cases the blunt leading edge of the initial airfoil is optimized into a sharp one, especially in ${\rm Kn}=0.01$ where the optimized airfoil has a quite long thin tip and looks just like we flip the initial airfoil left and right. As indicated in the figures, this sharp leading edge can effectively break the formation of the high-pressure area in front of the airfoil, leading to a significant reduction of the maximum pressure on the windward surface of the airfoil.
In ${\rm Kn}=0.01$ it can be seen clearly that the detached bow shock in front of the initial airfoil is transformed into two oblique shock waves at the leading edge of the optimized airfoil, making the high pressure occur only at the small tip and causing a big decrease of the wave drag. 
In ${\rm Kn}=0.5$ the shock waves are all dissipated out but we still can see a concave of the pressure contour in front of the optimized airfoil.
The detailed comparison of the shapes of the optimized airfoils is shown in Fig.~\ref{fig:case2_suf}, where we can see clearly that the optimized airfoil for ${\rm Kn}=0.01$ has a much thinner leading edge due to the stronger effect of the shock wave.

The drag coefficients for the initial and optimized airfoils are shown in Table \ref{tab:case2_cdeff}. For ${\rm Kn}=0.01$ there is a relatively big drag reduction of 13.36\% while for ${\rm Kn}=0.5$ the drag reduction is only 4.22\%. This is not surprising since we can see obvious changes of the shock wave for ${\rm Kn}=0.01$ before and after optimization, while for ${\rm Kn}=0.5$ the change of the flow field is relatively small. 
The computational cost is also shown in Table \ref{tab:case2_cdeff}. We adopt 80-core parallel computation (40 cores per node) with the CPU model ``\emph{Intel(R) Xeon(R) Gold 6148 CPU @ 2.40GHz}''. It can be seen that the optimizations generally converge in a dozen steps, due to the fast convergence of the quasi-Newton optimizer and the accuracy of present adjoint solver. In total, the optimizations take only around 5-7 minutes, showing the high efficiency of the present shape optimization method (as a method capable of handling the whole range of gas rarefaction).

\begin{table}[]
	\centering
	\caption{\label{tab:case2_cdeff}Optimization of the airfoil under supersonic flow: drag coefficients before/after optimization and the optimization efficiency. $C_{\rm d}$ is normalized by the far-field state $\frac{1}{2}\rho _\infty\bm u_\infty^2c$.}
	\vspace{0.3cm}
\begin{tabular}{ccccccccc}
\hline
\multirow{2}{*}{Kn} & \multicolumn{2}{c}{$C_{\rm d}$}                                      & \multirow{2}{*}{\begin{tabular}[c]{@{}c@{}}Drag\\ decrease\end{tabular}} & \multirow{2}{*}{\begin{tabular}[c]{@{}c@{}}Physical\\ mesh\end{tabular}} & \multirow{2}{*}{\begin{tabular}[c]{@{}c@{}}Velocity\\ mesh\end{tabular}} & \multirow{2}{*}{\begin{tabular}[c]{@{}c@{}}Parallel\\ cores\end{tabular}} & \multirow{2}{*}{\begin{tabular}[c]{@{}c@{}}Optim.\\ steps\end{tabular}} & \multirow{2}{*}{\begin{tabular}[c]{@{}c@{}}Time\\ (s)\end{tabular}} \\ \cline{2-3}
                      & \multicolumn{1}{l}{Initial} & \multicolumn{1}{l}{Optimized} &                                                                          &                                                                          &                                                                          &                                                                           &                                                                               &                                                                     \\ \hline
0.01              & 0.2615                      & 0.2266                        & 13.36\%                                                                     & 21666                                                                    & 28×28                                                                    & 80                                                                        & 25                                                                            & 311                                                                 \\
0.5                & 0.6909                      & 0.6617                        & 4.22\%                                                                                                                                         & 21666                                                                    & 1984                                                                    & 80                                                                        & 9                                                                             & 430                                                                 \\ \hline
\end{tabular}
\end{table}



\begin{figure}[t]
	\centering
	{\subfigure[${\rm Kn}=0.5$]{%
			\includegraphics[width=0.47\textwidth]{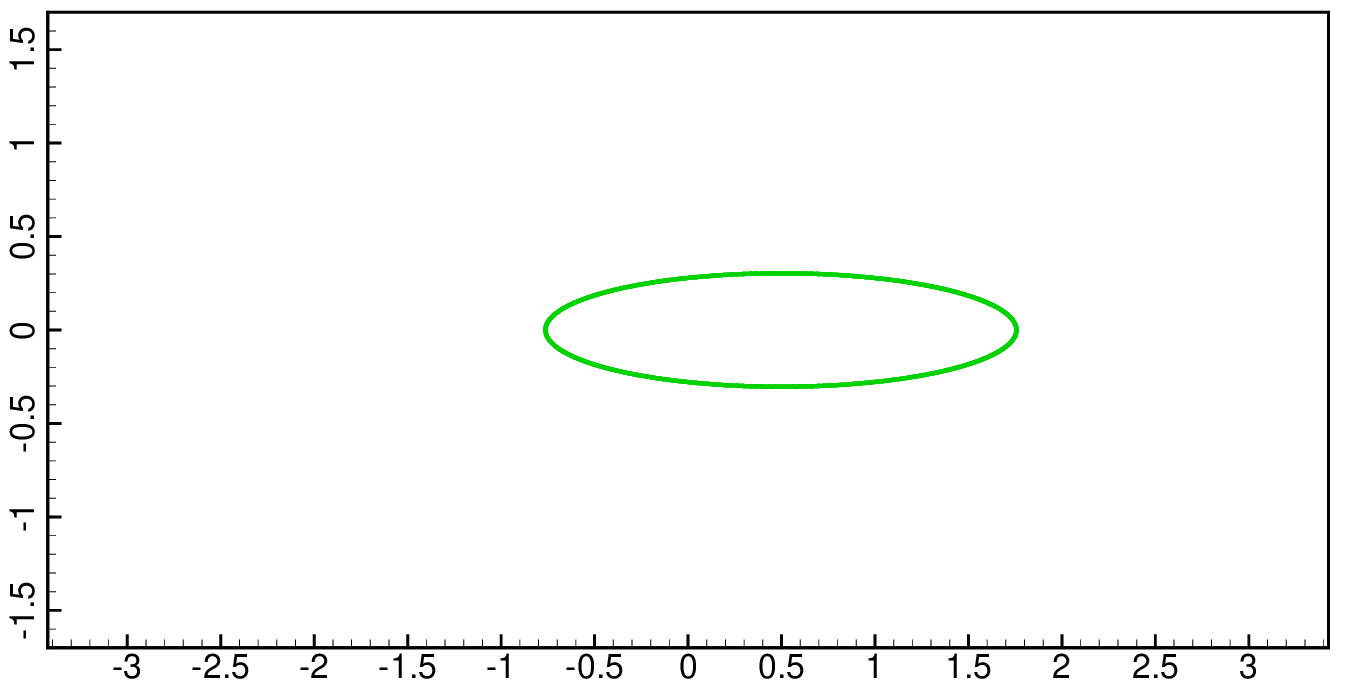}\hspace{0.02\textwidth}%
			\includegraphics[width=0.47\textwidth]{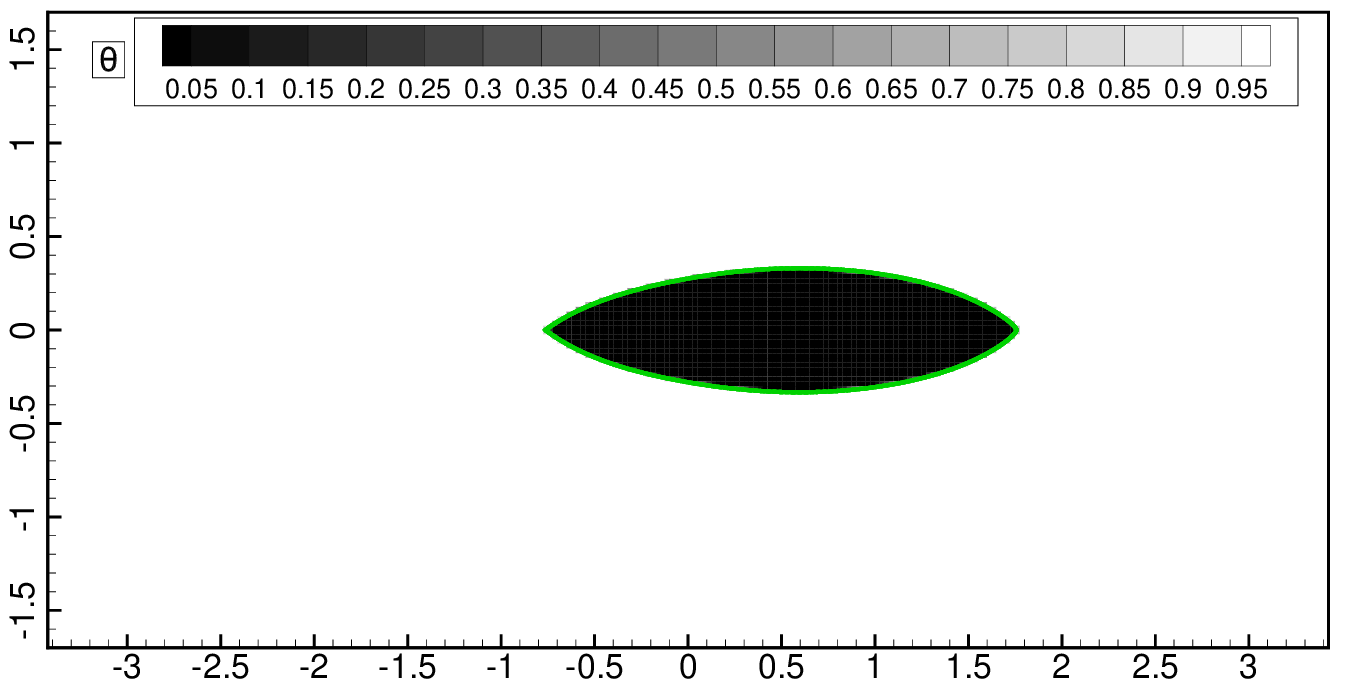}}}\\%
	{\subfigure[${\rm Re}=200$]{%
			\includegraphics[width=0.47\textwidth]{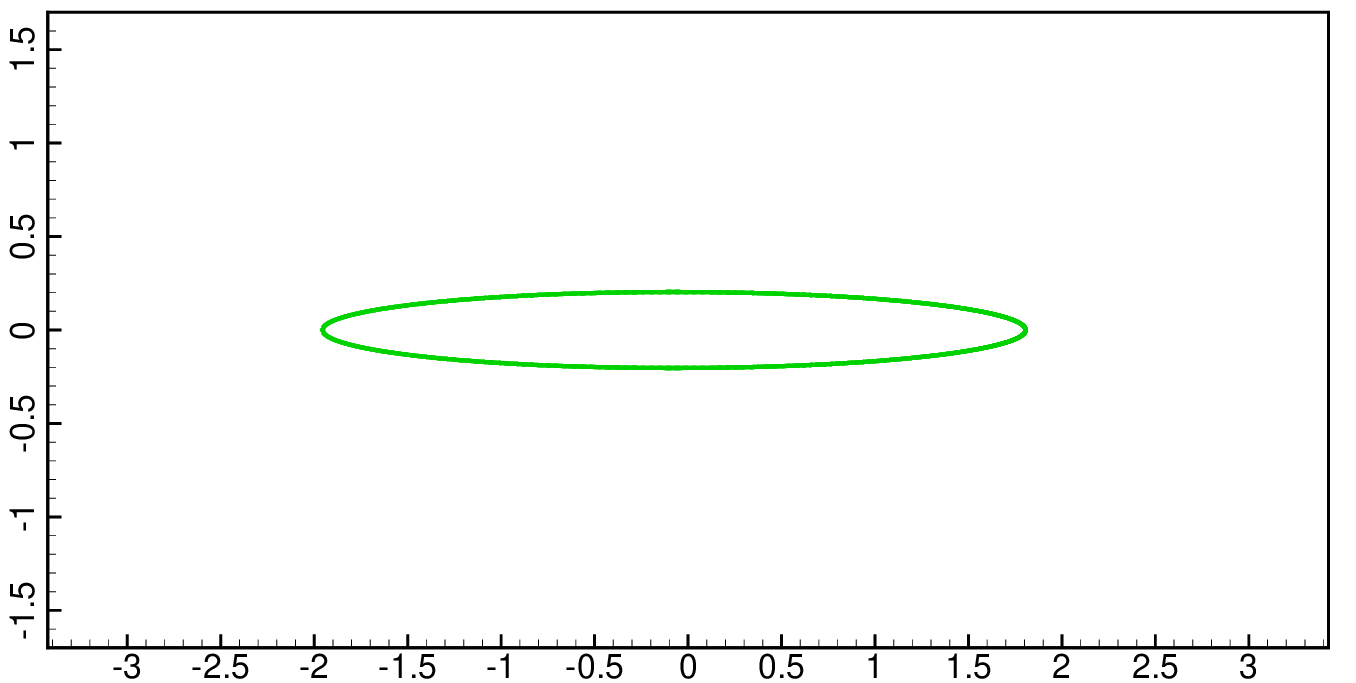}\hspace{0.02\textwidth}%
			\includegraphics[width=0.47\textwidth]{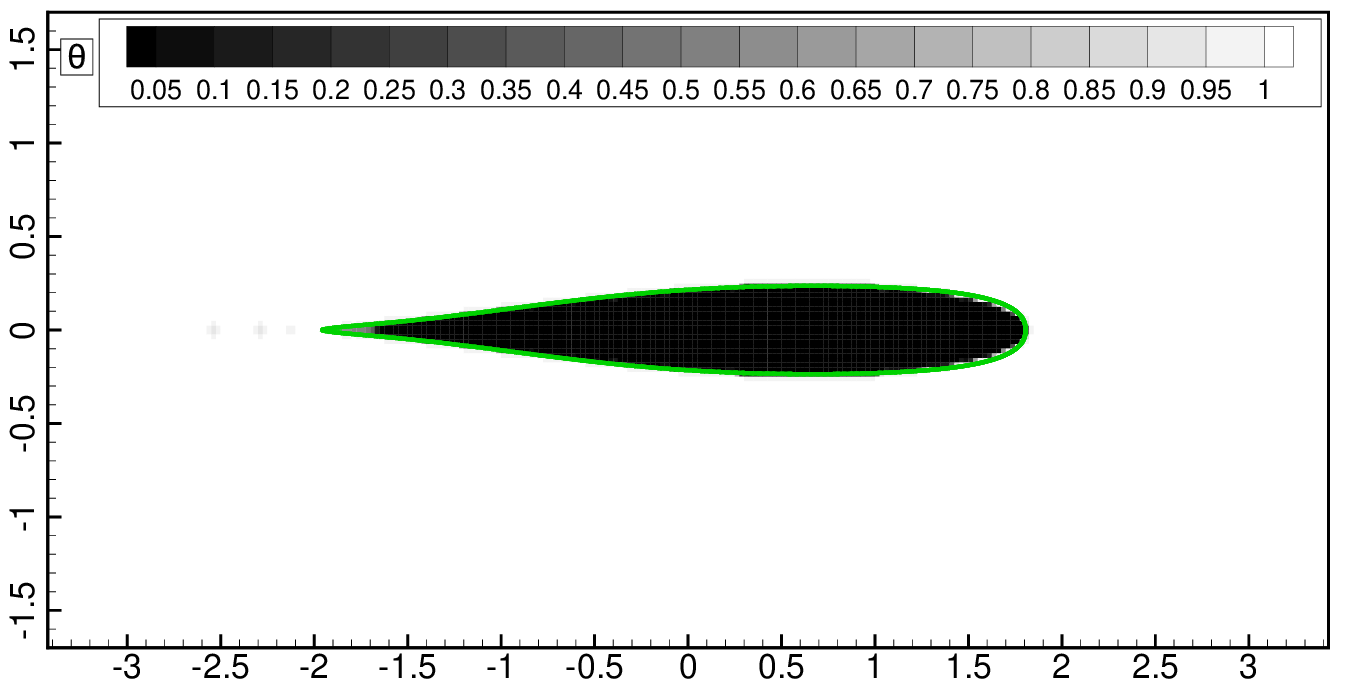}}}
	\caption{\label{fig:case2_topo}Optimization of the airfoil under supersonic flow: the comparison of the results calculated by the present shape optimization method (green lines) with the published results of Ref. \cite{yuan2024design} obtained by the topology optimization method (grayscale fields). The left column is the initial setup for the present method. Kn and Re are defined by a reference length of 2 (completely same condition with Ref. \cite{yuan2024design}).}
\end{figure}

Similar to Section \ref{sec:case1}, for the test cases of this section we have also compared the results with those obtained by the topology optimization in \cite{yuan2024design}. As explained in Section \ref{sec:case1}, here we start the optimization from the initial shape of an elliptic cylinder of the same area-length ratio $V_{\min}/c^2$ with that of the optimized airfoil in \cite{yuan2024design}. The initial shapes and the optimal results are shown in Fig.~\ref{fig:case2_topo}. It is shown that the two optimization methods, which have totally different styles of description for the airfoil geometry, yield almost the same optimal airfoil shapes, verifying the present method from another perspective.

\section{Conclusions}\label{sec:conc}

In this paper, a shape optimization method has been proposed, applicable to design problems involving both rarefied and continuum gas flows. The gas flow is governed by the Boltzmann-BGK model equation, with the diffuse boundary condition imposed on the gas-solid surface, whose geometry is represented by the CST parameterization. The sensitivity with respect to the geometry parameters of the solid boundary has been calculated using a blend of the continuous adjoint and the discrete adjoint approaches, where we first obtained the adjoint variable through continuous adjoint analysis and then computed the sensitivity in a manner akin to the discrete adjoint approach. To ensure the accuracy and efficiency of the adjoint method across different flow regimes, both the primal and adjoint kinetic equations were solved using implicit multiscale gas-kinetic schemes with the prediction acceleration technique. For the optimizer, we employed the sequential least-squares quadratic programming algorithm based on the quasi-Newton method with the BFGS formula. The mesh deformation due to shape optimization was handled by the radial-basis-function interpolation method.

To validate the present method, we first investigated the sensitivity calculated by our method in a test case involving flow over an elliptic cylinder inside a channel. It has been found that the sensitivity with respect to the coordinates of the mesh points defining the solid surface suffered from severe oscillations, which seemed due to the discretization of the molecular velocity space for solving the gas-kinetic governing equation. This issue has been well resolved by the implementation of the CST parameterization, which acted as a low-pass filter and thoroughly eliminated the oscillations. Nonetheless, the sensitivities obtained by our method and the finite difference method agreed well with each other.

We then verified the present method through optimizations of the airfoil shape for drag reduction. Two sets of cases were carried out: the airfoil inside a channel and the airfoil under a Mach number of 2 supersonic flow, covering gas rarefaction from Kn = 10 to Kn = 0.001, i.e., from free-molecular to continuum regimes. It has been demonstrated that the airfoil has different optimal shapes for different degrees of gas rarefaction, and the present method could provide optimal solutions within a dozen optimization steps with a time cost of 5 to 20 minutes (parallel computation with 40 to 160 cores). The results obtained have been compared with those from the topology optimization method in previous work \cite{yuan2024design}, and good consistency was achieved.


In conclusion, the present method has demonstrated its effectiveness and efficiency in addressing shape optimization problems for gas flows spanning from rarefied to continuum conditions. Extending our method to 3D problems is straightforward: the primary task is to establish the derivatives of the surface elements' attributes with respect to the nodes' coordinates (i.e., the 3D equivalent of equation \eqref{eqn:sens_nod}), which involves only differentiating geometric relationships. Once this is done, the rest of the method can be seamlessly extended to the 3D case. Looking ahead, we plan to integrate our optimization method with a cutting-edge 3D multiscale gas flow solver \cite{zhang2024efficient} to create a robust shape optimization tool for 3D design challenges that encompass both rarefied and continuum gas flows.

\section*{Acknowledgments}
The authors thank Prof. Steven G. Johnson (Massachusetts Institute of Technology) for the implementation of SLSQP algorithm in the NLopt library \cite{johnson2007nLopt}. This work is supported by the National Natural Science Foundation of China (12402388). The authors acknowledge the computing resources from the Center for Computational Science and Engineering at the Southern University of Science and Technology.



%


\bibliographystyle{elsarticle-num} 
\bibliography{2024_adjoint}

\end{document}